\documentclass[10pt]{article}
\usepackage{times}

\usepackage{amstext}
\usepackage{amscd}
\usepackage{longtable} 
  \usepackage{colortbl}
 \usepackage{arydshln}
\usepackage{bm}
\usepackage{natbib}
\usepackage{url}
\usepackage{here}
\usepackage{subeqnarray}
\usepackage{indentfirst}
\usepackage{subfigure}
\usepackage{appendix}
\usepackage{rotating}
\usepackage{multirow}
\usepackage{fancyhdr}
\usepackage{latexsym}
\usepackage{verbatim}
\usepackage{setspace}
\usepackage{amsmath,amsfonts}
\usepackage{graphicx,psfrag}
\usepackage{enumerate}
\usepackage{color}
\usepackage{lineno}
\usepackage{listings}

\definecolor{codegreen}{rgb}{0,0.6,0}
\definecolor{codegray}{rgb}{0.5,0.5,0.5}
\definecolor{codepurple}{rgb}{0.58,0,0.82}
\definecolor{backcolour}{rgb}{0.95,0.95,0.92}

\definecolor{blue}{rgb}{0.0, 0.2, 0.6}
\definecolor{brown}{rgb}{0.65, 0.16, 0.16}
 
\lstdefinestyle{mystyle}{
    backgroundcolor=\color{backcolour},   
    commentstyle=\color{codegreen},
    keywordstyle=\color{magenta},
    numberstyle=\tiny\color{codegray},
    stringstyle=\color{codepurple},
    basicstyle=\footnotesize,
    breakatwhitespace=false,         
    breaklines=true,                 
    captionpos=b,                    
    keepspaces=true,                 
    numbers=left,                    
    numbersep=5pt,                  
    showspaces=false,                
    showstringspaces=false,
    showtabs=true,                  
    tabsize=2
}
 
\def \build#1#2#3{\mathrel{\mathop{#1}\limits^{#2}_{#3}}}

\def \diag {\mbox{diag}}

\numberwithin{equation}{section}

\begin{document}

\title{\LARGE \sc Zero-adjusted Birnbaum-Saunders regression model}


\author{{\bf  \normalsize \small Vera Tomazella${}^{1}$,\,Juv\^encio S. Nobre${}^2$ ,\,Gustavo H.A. Pereira${}^1$ \, and \,Manoel Santos--Neto${}^{1,3}$\thanks{Corresponding author:  Email: mn.neco@gmail.com URL: www.santosnetoce.com.br}} \\
\small{$^1$Departamento de Estat\'istica, Universidade Federal de S\~ao Carlos, Brazil}\\
\small{$^2$Departamento de Estat\'istica e Matem\'atica Aplicada, Universidade Federal do Cear\'a, Brazil}\\
\small{$^3$Departamento de Estat\'istica, Universidade Federal de Campina Grande, Brazil}
}

\date{}

\maketitle

\begin{abstract}
In this paper we introduce the zero-adjusted Birnbaum-Saunders regression model. This new model generalizes at least seven Birnbaum-Saunders regression models. The idea of this modeling is mixing a degenerate distribution at zero with a Birnbaum-Saunders distribution. Besides the capacity to account for excess zeros, the zero-adjusted Birnbaum-Saunders distribution additionally produces an attractive modeling structure to right-skewed data. In this model, the mean and precision parameter of the Birnbaum-Saunders distribution and the probability of zeros can be related to linear and/or non-linear predictors through link functions. We derive a type of residual to perform diagnostic analysis and a perturbation scheme for identifying those observations that exert unusual influence on the estimation process. Finally, two applications to real data show the potential of the model.
\paragraph{Key Words:} Birnbaum-Saunders distribution; Nonlinear Regression Models; Reparameterization; Zero-adjusted.
\end{abstract}

\section{Introduction}

Although distributions for zero-inflated count data have attracted the most attention, the study of continuous data with zero inflation has, in fact, a longer history. Concerns about statistical analysis of zero-inflated data were first identified in \cite{Aitchinson1955} that proposed a mixture of zero with a lognormal model resulting in the well-known \textit{delta distribution}.  In the literature, there are various examples of inflated distributions, for example, \cite{feu1979}, \cite{Farewell:1986aa}, \cite{Meeker:1987aa}, \cite{Lambert:1992aa}, \cite{Shankar:1997aa}, \cite{IWASAKI:2009aa}, \cite{Heller:aa}, \cite{Ospina:2008aa}, \cite{Tong:2013aa}, \cite{Pereira:2012aa}, \cite{Ospina:2012aa}, among others.  \cite{Tu:2014aa} briefly reviewed the concept of zero inflation and surveyed existing analytical methods for zero-inflated data.

When a set of Poisson (PO) distributed count data includes more zeros than the amount expected under a PO probability mass function (PMF), it is declared to be ``zero-inflated". Beyond the ``zero-inflated" term the idea of joining a degenerate model at zero with a continuous model can be named ``zero-adjusted" or ``zero-augmented"~\citep{Galvis:2014aa}. In this article, we use the term ``zero-adjusted", as in~\cite{Heller:aa} and \cite{lscb:15}.

If positive continuous data include a point-mass at zero, it is necessary to contemplate whether the zeros encountered represent censored observations due to a limit of detection, true zeros due to actual observations not due to censoring, or a mixture of true zeros and censored values. 
In medical spending cases, the zero spending values noted are true zeros because it was noted directly that the individuals or household had no medical claims during the study term~\citep[][]{tuzhou1999, zhoutu1999, duanetal1983}.

Conceptually, a true zero is different from a censored zero. A true zero typically symbolizes something not happening, whereas a censored zero only means that its occurrence was below a certain threshold. To deal with such situations, conventional likelihood methods have been proposed to test for the presence of censoring~\citep[][]{berk2002}.
Such observations usually require a different strategy, and will not be presented in this article.

Zero-adjusted models assume that the zero value is really observed. In several areas,  there are many examples of these distributions:  in car insurance studies, the total claim amount reported to a given contract is often equal to 0, if no claims have been filed against the insurer, but may also be strictly positive if one or diverse accidents occurred; in microbiology, such non-negative data could happen from assays, virus titers, or metabolomic and proteomic data~\citep[][]{taylor2009};  finally, in economic studies, the amount an individual or household spends on a determined category during the study period is semicontinuous~\citep[see, for example,][]{tuzhou1999,zhoutu1999}.

In recent years, the Birnbaum-Saunders (BS) distribution~\citep{bs:69b,bs:69a}  has received much attention, due to its theoretical formulation associated with cumulative damage processes, its relation to the normal distribution, and its mathematical properties.  In \cite{scla:12}  several parameterizations for the BS distribution by using different arguments was proposed. One of them indexes the BS distribution by its mean and precision, and it is named the reparameterized BS (RBS) distribution. \cite{lscb:14,lscb:15}, \cite{sclb:14, sclb:16} and \cite{llst:17} showed that the RBS model is useful in frameworks for which the original parameterization is limited.

\cite{lscb:15} proposed a methodology for inventory logistics that allows demand data that have zeros to be modeled by means of a new discrete-continuous mixture distribution, which is constructed by using a probability mass at zero and a continuous component related to the RBS distribution. They named this new class of models as the zero-adjusted RBS (ZARBS) distribution. Based on this model we propose a general class of regression models for modeling semicontinuous data. For simplicity, the RBS distribution will be called BS distribution and the corresponding regression model that will be proposed by zero-adjusted Birnbaum-Saunders (ZABS) regression model.

In this article, we shall allow the mean and the precision parameter of the BS distribution and the probability of a point mass at 0 to be related to linear or non-linear predictors through link functions.  This model generalizes the models proposed by \cite{lscb:14} and \cite{sclb:16}. Some particular models that usually arise in the practical application are: (i) the linear BS regression model (BSRM); (ii) the linear BSRM with varying precision; (iii) the nonlinear BS regression model (NBSRM); (iv) the NBSRM with linear precision covariates; (v) the linear ZABS regression model; (vi) the linear ZABS regression model with varying precision. Additionally, non-constant response variances are naturally adjusted once the variance of the response variable is a function of the covariates.

Apart from this introduction, the paper is organized as follows. Section~\ref{sec:2} presents a general class of ZABS regression models. Section \ref{sec3} is devoted to diagnostic analysis.  Section \ref{sec4} contains two applications using real data and concluding remarks are given in Section \ref{sec5}. Some technical details are collected in an appendix.

\section{ZABS model} \label{sec:2}

The probability density function (PDF) of the BS distribution~\citep[]{scla:12} with strictly positive values of the 
precision parameter $\sigma$ and mean parameter $\mu$, is given by
\begin{equation}
f_T(t|\mu,\sigma)=\frac{\exp\left(\sigma/2\right)\sqrt{\sigma+1}}{4\sqrt{\pi\mu}\,t^{3/2}}
\left[t+\frac{\sigma \mu}{\sigma+1}\right] \exp\left(-\frac{\sigma}{4}
\left[\frac{\{\sigma+1\}t}{\sigma\mu}+\frac{\sigma\mu}{\{\sigma+1\}t}\right]\right).
\label{eq:i2}
\end{equation}
The mean and variance of the BS distribution are given, respectively,  by
\[
\textrm{E}[T] = \mu \quad \text{and} \quad \textrm{Var}[T] = \frac{(2\,\sigma+5)}{(\sigma+1)^2}.
\]

We say that a random variable $Y$ follows a ZABS distribution with parameters $\mu, \sigma$ and $\nu$, denoted by ZABS$(\mu,\sigma,\nu)$, if the distribution of $Y$ admits the following PDF   
\begin{eqnarray*}
 f_{Y}(y|\mu,\sigma,\nu)  =  \left\{ [1-\nu]
\frac{\exp(\frac{\sigma}{2})\sqrt{\sigma+1}}{4\,y^{3/2}\,\sqrt{\pi\mu}}\left[y+\frac{\sigma
\mu}{\sigma+1} \right] \exp\left(-\frac{\sigma}{4}\left[\frac{\{\sigma+1\} y}{\sigma\mu}+
\frac{\sigma\mu}{\{\sigma+1\} y}\right]\right)\right\}^{1-\mathbb{I}(y = 0) } \times 
\nu^{\mathbb{I}(y=0)},
\end{eqnarray*}
\noindent
with, $\sigma >0$, $\mu >0$ and $0<\nu<1$. We have that the mean and variance of $Y \sim \text{ZABS}(\mu,\sigma, \nu)$ are, respectively, given by
$
\textrm{E}[Y] = (1-\nu)\, \mu \quad \mbox{and} \quad \textrm{Var}[Y] = (1-\nu)\mu^2 \left\{ \nu +
\textrm{CV}[T]^2\right\},
\label{eq:11}
$
where $\textrm{CV}[T] = \frac{\sqrt{(2\,\sigma+5)}}{(\sigma+1)}$.
The log-PDF of the ZARBS distribution is given by
\begin{equation}
 \ell(\mu,\sigma,\nu) = \log(\nu)\, \mathbb{I}(y=0) + \log(1-\nu)\,[1-\mathbb{I}(y=0)] + [1-\mathbb{I}(y=0)]\,\log(f_T(y)).
  \label{ld}
\end{equation}
Let $Y_1,\ldots,Y_n$ be a random sample from $Y \sim \text{ZABS}(\mu,\sigma, \nu)$.
Then, the corresponding likelihood function for $\bm \theta = [\mu,\sigma,\nu]^\top$ is given by
\begin{equation}
\mathrm{L}(\mu,\sigma, \nu) = \prod_{i=1}^{n}f_Y(y_i|\mu,\sigma, \nu) =
\nu^{n_0}[1-\nu]^{n - n_0} \prod_{i=1}^{n}f_{T}(y_i|\mu,\sigma)^{1-\mathbb{I}(y_i= 0)},\label{eq:12}
\end{equation}
where $n_0 = \sum_{i=1}^{n} \mathbb{I}(y_i= 0)$ is the number of observations equal to zero.
Hence, the corresponding log-likelihood function obtained from \eqref{eq:12} can be
expressed as $\ell({\bm \theta}) = \ell(\nu) + \ell(\mu,\sigma)$, where
\begin{align}\label{eq:13}
\ell(\nu) &= n_0 \log(\nu) + [n-n_0]\log(1-\nu),\\
\ell(\mu,\sigma) &= [n-n_0]c(\mu,\sigma) -
\frac{3}{2}\sum_{y_i>0}\log(y_i)
-\frac{[\sigma+1]}{4\mu}\sum_{y_i>0}y_i  - \sum_{y_i>0}\frac{\mu
\sigma^2}{4[\sigma+1]y_i}+
\sum_{y_i>0}\log\left(y_i + \frac{\mu \sigma}{[\sigma+1]}\right),\nonumber
\end{align}

\noindent with $c(\mu,\sigma) =-[1/2]\log(16\pi) + [\sigma/2]
-[1/2]\log(\mu) + [1/2]\log(\sigma+1)$.

Let $Y_1,\ldots,Y_n$ be a random sample such that each $Y_i \sim \text{ZABS}(\mu_i, \sigma_i ,\nu_i)$. Suppose the mean, precision and mixture parameters of $Y_i$ satisfy the following functional relations:  
\begin{equation*}
g_1(\mu_i) = \eta_i = f_1(\mathbf{x}_i;{\bm \beta}), \quad g_2(\sigma_i) = \tau_i = f_2(\mathbf{z}_i;{\bm \alpha}), \quad \mbox{and} \quad g_3(\nu_i) = \xi_i 
= f_3(\mathbf{w}_i;{\bm \gamma}),
\label{est}
\end{equation*}
for $\quad i=1,\ldots,n$, where ${\bm \beta}=[\beta_1,\ldots,\beta_{p}]^\top$, ${\bm \alpha} = [\alpha_1,\dots,\alpha_q]^\top$ and 
${\bm \gamma} = [\gamma_1,\dots,\gamma_r]^\top$ are vectors of unknown parameters to be estimated, for $p+q+r<n$,
${\bm \eta} = [\eta_1,\dots,\eta_n]^\top$, ${\bm \tau} = [\tau_1,\dots,\tau_n]^\top$ and
${\bm \xi} = [\xi_1,\dots,\xi_n]^\top$ are predictor vectors, and $f_j(\cdot;\cdot), j=1,2,3$ are linear or nonlinear twice 
continuously differentiable functions in the second argument, such that the derivative matrices $\widetilde{\mathbf{X}} = \partial \bm \eta/\partial \bm \beta$ , $\widetilde{\mathbf{Z}} = \partial \bm \tau/\partial \bm \alpha$ and $\widetilde{\mathbf{W}} = \partial \bm \xi/\partial \bm \gamma$ are full ranks for all $\bm \beta$, $\bm \alpha$ and $\bm \gamma$.  Moreover, $\tilde{\mathbf{x}}_i=[\tilde{\mathrm{x}}_{i1},\ldots,\tilde{\mathrm{x}}_{ip_1}]^{\top}$, 
$\tilde{\mathbf{z}}_i=[\tilde{\mathrm{z}}_{i1},\ldots, \tilde{\mathrm{z}}_{ip_2}]^{\top}$ and $\tilde{\mathbf{w}}_i=[\tilde{\mathrm{w}}_{i1},\ldots, \tilde{\mathrm{w}}_{ip_3}]^{\top}$ 
are vectors that contain the values of $p_1, p_2$ and $p_3$ explanatory variables, respectively. 
In this model, the link functions $g_j \mbox{: } \mathbb{R}^{+} \rightarrow \mathbb{R} ,  j=1,2$ 
are strictly monotone, positive, and at least twice differentiable and $g_3 \mbox{: } (0,1) \rightarrow \mathbb{R}$ 
is strictly monotonic and twice differentiable.

The log-likelihood function for this model is given by $\ell(\mu_i,\sigma_i, \nu_i) = \ell(\nu_i) + \ell(\mu_i,\sigma_i)$, with
the expressions of $\ell(\nu_i) $ and $ \ell(\mu_i,\sigma_i)$ given as \eqref{eq:13}. Here, $\mu_i = g_1^{-1}(\eta_i)$, 
$\sigma_i = g_2^{-1}(\tau_i)$ and $\nu_i = g_3^{-1}(\xi_i)$. We note that $\ell(\mu_i,\sigma_i)$ is the log-likelihood function
for $[{\bm \beta},{\bm \alpha}]^\top$ in a nonlinear BS regression model with varying precision. Furthermore, $\ell(\nu_i)$
represents the log-likelihood function of a regression model for binary responses. Therefore, the maximum likelihood 
estimations (MLE's) for this model can be accomplished by separately fitting a nonlinear binomial regression model and 
a nonlinear BS regression model with varying precision.

Differentiation of the log-likelihood function with respect to the unknown parameters provides the score function, which is given by
\(
\mathbf{U}_{\mathbf{\theta}} = [\mathbf{U}_{{\bm \beta}}^\top, \mathbf{U}_{{\bm \alpha}}^\top, \mathbf{U}_{{\bm \gamma}}^\top]^\top,
\)
where ${\bm \theta} = [{\bm \beta}^\top,{\bm \alpha}^\top,{\bm \gamma}^\top]^\top$ and
\(
\mathbf{U}_{{\bm \beta}} = \widetilde{\mathbf{X}}^\top\,\mathbf{A}_\kappa\,[\mathbf{y}^{*} - {\bm \mu}^{*}], \quad
\mathbf{U}_{{\bm \alpha}} = \widetilde{\mathbf{Z}}^\top\,\mathbf{B}_\kappa\,[\mathbf{y}^{\bullet} - {\bm \sigma}^{\bullet}], \quad \text{and} \quad
\mathbf{U}_{{\bm \gamma}}= \widetilde{\mathbf{W}}^\top\,\mathbf{C}\,[\mathbf{y}^{\circ} - {\bm \nu}^{\circ}], 
\)
and this results are shown in~\ref{ap1}. Thus, the Fisher information matrix is given by
\begin{equation*}
\arraycolsep=2.5pt\def\arraystretch{2.2}
\setlength{\dashlinegap}{2pt}
\mathbf{i}_{\bm \theta}= \left[
\begin{array}{ccc:c}
\widetilde{\mathbf{X}}^\top\, \mathbf{V} \widetilde{\mathbf{X}} & &\widetilde{\mathbf{X}}^\top\, \mathbf{S} \widetilde{\mathbf{Z}} & \mathbf{0} \\ 
\widetilde{\mathbf{Z}}^\top\,  \mathbf{S} \widetilde{\mathbf{X}}&& \widetilde{\mathbf{Z}}^\top\,  \mathbf{U} \widetilde{\mathbf{Z}} & \mathbf{0}\\  \hdashline
 {\bm 0} && {\bm 0} & \widetilde{\mathbf{W}}^\top \mathbf{Q} \widetilde{\mathbf{W}} 
\end{array}
\right],
\end{equation*}
and the matrix elements are presented in \eqref{inf}. Under general regularity conditions, the ML estimators of $\bm \theta$ and $\mathbf{i}_{\bm \theta}$, $\widehat{{\bm \theta}}$ and $\mathbf{i}_{\widehat{{\bm \theta}}}$ say, respectively, are consistent. Suppose that $\mathbf{j}_{\bm \theta}=\lim \limits_{n \to \infty}[1/n]\mathbf{i}_{\bm \theta}$ exists and is non-singular, then
\[
\sqrt{n} [\widehat{{\bm \theta}} -{\bm \theta}] \build{\to}{d}{} \text{N}_{p+q+r}({\bm 0}, \mathbf{j}_{\bm \theta}^{-1}),
\]
 where $\build{\to}{d}{}$ denotes convergence in distribution to. Thus, if $\theta_k$ denotes the $k$th element of ${\bm \theta}$, $[\widehat{\theta}_k - \theta_k]/\sqrt{\mathrm{v}_{k}} \build{\to}{d}{} \text{N}(0,1)$, where $ \mathrm{v}_k$ is $k$th diagonal element of the matrix $\mathbf{i}_{\bm \theta}^{-1}$ defined in \eqref{invF}, for $k=1,\ldots,p+q+r$. 
  Then, asymptotic confidence interval for $\theta_k$ and asymptotic coverage probability $100[1-\upsilon]\%$, is given by $\widehat{\theta}_k \pm z_{1-\upsilon/2}\cdot\mathrm{v_k}^{1/2}$ where $z_\upsilon$ denotes the $\upsilon$th quantile of the $\text{N}(0,1)$ distribution.
  The confidence intervals and the confidence region can be used to test hypotheses about the parameters. The asymptotic distribution of the maximum likelihood estimator of any differentiable function of ${\bm \theta}$ can be obtained using the delta method.


Theoretical results of this paper have been implemented in the \textbf{R}~\citep{r:17} software. In special, the \textbf{RBS} package provides functions for fitting Birnbaum-Saunders regression models using the \textbf{gamlss} package. The current version can be downloaded from GitHub via

\begin{lstlisting}[language=R]
devtools::install_github("santosneto/RBS")
\end{lstlisting}

This package contain a collection of utilities for analyzing data from RBS and ZARBS distributions. For example, the functions \textbf{dRBS}, \textbf{pRBS}, \textbf{qRBS} and \textbf{rBS} define the density, distribution function, quantile function and random generation for the RBS distribution. Other functions are: {\bf dZARBS}, {\bf rZARBS}, {\bf qZARBS}, {\bf pZARBS}, \textbf{envelope} and \textbf{envelope.ZARBS} .  

\section{Diagnostic analysis}\label{sec3}

In this section, we will present a residual and a perturbation scheme for identifying those observations that exert unusual influence on the estimation process and consequently on the parameter estimates. 

\subsection{Residuals}
To assess goodness of fit and departures from the assumptions of the ZABS regression model, we propose the randomized quantile residual \citep[][]{ds:96}. It is a randomized version of the \cite{cs:68} residual and is given by
\begin{equation}\label{rq}
r_i^q  = \left\{ 
\begin{array}{cc}
\Phi^{-1}(f_i), & \text{if} \quad y_i >0; \\
\Phi^{-1}(u_i), & \text{if} \quad y_i=0,
\end{array}
\right.
\end{equation}
where $\Phi^{-1}(\cdot)$ is the inverse function of the standard normal cumulative distribution function (CDF),
$f_i = F(y_i, \hat \mu_i, \hat \sigma_i, \hat \nu_i)$ is the estimated value using the CDF of the ZABS distribution 
and
$u_i$ is the observed value of a random variable $U_i$ with uniform distribution in $(0,\hat \nu_i)$, where $\hat \nu = \widehat{\textrm{P}}(Y_i = 0)$. 
 If the model is correctly specified, then $r_i^q$ is approximately normally distributed.   

The residuals versus the index of the observations should show neither detectable pattern. A trend in the plot of some residual versus the predictors or the fitted values may be indicative of link function misspecification. Furthermore, normal probability plots with simulated envelopes are a helpful diagnostic instrument \citep[][]{atkinson1985}.

\subsection{Local Influence}

The concept of local influence was introduced by \cite{c:86}. This approach based on normal curvature is an important diagnostic
tool for assessing local influence of minor perturbations to a statistical model.  \cite{c:86} suggested that the influence of the perturbation $\omega$ could be studied using the likelihood displacement function defined as 
$\textrm{LD}= 2[{\ell}(\widehat{\bm{\theta}}) - {\ell}(\widehat{\bm
{\theta}}_\omega)],$ where ${\ell}(\widehat{\bm{\theta}})$ and  ${\ell}(\widehat{\bm
{\theta}}_\omega)$ are log-likelihood functions of the postulate and perturbed models, respectively. 
The normal curvature for $\bm{\theta}$ in the direction vector $\mathbf
{d}$, with $||\mathbf{d}||=1$, is expressed as $\mathrm{C}_{\mathbf{d}} = 2|
\mathbf{d}^\top{\bm{\Delta}}^\top\,\mathbf{h}_{\bm \theta}^{-1}\,{\bm{\Delta}}\,\mathbf{d}|$, where ${\bm{\Delta}}=
\partial^2{\ell}(\bm{\theta}|{\bm{\omega}})/\partial\bm \theta\partial\bm \omega$ is the matrix of perturbations.  A local influence diagnostic tool is generally based on index plots. The index plot of the direction of maximum curvature, $\mathbf{d}_{\max}$, corresponding to the maximum eigenvalue of
\(
\mathbf{F}_{\bm{\theta}} = {\bm{\Delta}}^\top\, \mathbf{h}_{\bm \theta}^{-1}\, {\bm{\Delta}},
\)
$\mathrm{C}_{\mathbf{d}_{\max}}$, evaluated at $\bm{\theta} =
\widehat{\bm{\theta}}$ and $\bm \omega = \bm \omega_0$, can contain valuable diagnostic information. Also, the index plot of
$\mathrm{C}_i= 2 |f_{ii}|$, where $f_{ii}$ is the $i$th diagonal
element of
$\mathbf{F}_{\bm \theta}$, can be used as a 
diagnostic technique to evaluate the existence of influential observations.
Those cases when
$\widehat{\mathrm{C}}_i
> 2\cdot\overline{\mathrm{C}}$, where $\overline
{\mathrm{C}} =
\sum^{n}_{i=1}\widehat{\mathrm{C}}_i/n$, are considered as
potentially influential.

\subsubsection*{Case-weights perturbation}

We define a perturbation $\bm \omega = [\omega_1,\ldots,\omega_n]^\top$, $\omega_i \geq 0$, to perturb
the contribution of each case to the log-likelihood. In the ZABS regression model, we have that the form of $\mathbf{\Delta}_{\bm \beta}$, $\mathbf{\Delta}_{\bm \alpha}$ and $\mathbf{\Delta}_{\bm \gamma}$ are, respectively, given by 
$
\mathbf{\Delta}_{\bm \beta}= \widetilde{\mathbf{X}}^{\top}\,\mathbf{A}_\kappa\, \left[(y_i^{*} - \mu_i^{*})\,\delta^{n}_{ii}\right]$, 
$\mathbf{\Delta}_{\bm \alpha}= \widetilde{\mathbf{Z}}^{\top}\,\mathbf{B}_\kappa\, \left[(y_i^{\bullet}-\sigma_i^{\bullet})\, \delta^{n}_{ii}\right]$ 
and
$ 
\mathbf{\Delta}_{\bm \gamma}= \widetilde{\mathbf{W}}^{\top}\,\mathbf{C}\, \left[(y_i^{\circ} - \nu_i^{\circ})\,\delta^{n}_{ii}\right]$, evaluated at $\widehat{ \bm \theta}$ and $\bm \omega_0 =  {\bm 1}_n$ ,that is, an $n \times 1$ column vector with all elements equal to 1 and $\delta^{n}_{ii}$ is a diagonal matrix $n \times n$.


\section{Applications} \label{sec4}
To illustrate the methodology developed in the previous sections, we will consider two datasets. First, we will consider an application with a particular case of the ZABS nonlinear regression model where the $f_1(\cdot)$ and $f_3(\cdot)$ are linear functions and the parameter $\sigma$ is constant. Second, we will consider an application with a particular case of the ZABS nonlinear regression model where the parameter $\nu$ is stochastically equal to zero and the parameter $\sigma$ is constant.
\subsection{ZABS regression model}
In this application the data set refers to the study carried out by the Institute of Biomedical Sciences, University of S\~ao Paulo, that studied the fumonosin production by Fusarium verticilliodes in corn grains~\citep[]{Rocha:2017aa}. Here, we study the relation between the variables Fusarium verticillioides ($x_1$), Time ($x_2$), Precipitation ($x_3$) and Water activity ($x_4$) and the fumonisin B2 (FB2) production ($y$). 

\begin{figure}[H]
 \centering
\psfrag{Corr}[c][c]{Corr}
\psfrag{0.283}[c][c]{$0.283$}
\psfrag{0.562}[c][c]{$0.562$}
\psfrag{0.249}[c][c]{$0.249$}
 \psfrag{-0.236}[c][c]{$-0.236$}
\psfrag{0.452}[c][c]{$0.452$}
\psfrag{-0.285}[c][c]{$-0.285$}
\psfrag{-0.253}[c][c]{$-0.253$}
\psfrag{-0.416}[c][c]{$-0.416$}
\psfrag{-0.593}[c][c]{$-0.593$}
 \psfrag{-0.0498}[c][c]{$-0.0498$}
 \psfrag{y}[c][c]{$y$}
 \psfrag{x1}[c][c]{$x_1$}
 \psfrag{x2}[c][c]{$x_2$}
 \psfrag{x3}[c][c]{$x_3$}
 \psfrag{x4}[c][c]{$x_4$}
 \psfrag{0}[c][c][0.5]{$0$}
 \psfrag{1}[c][c][0.5]{$1$}
 \psfrag{2}[c][c][0.5]{$2$}
 \psfrag{3}[c][c][0.5]{$3$}
 \psfrag{4}[c][c][0.5]{$4$}
 \psfrag{5}[c][c][0.5]{$5$}
 \psfrag{10}[c][c][0.5]{$10$}
 \psfrag{15}[c][c][0.5]{$15$}
 \psfrag{20}[c][c][0.5]{$20$}
 \psfrag{25}[c][c][0.5]{$25$}
 \psfrag{30}[c][c][0.5]{$30$}
 \psfrag{40}[c][c][0.5]{$40$}
 \psfrag{50}[c][c][0.5]{$50$}
 \psfrag{60}[c][c][0.5]{$60$}
 \psfrag{75}[c][c][0.5]{$75$}
 \psfrag{100}[c][c][0.5]{$100$}
 \psfrag{0.98}[c][c][0.5]{$0.98$}
 \psfrag{0.96}[c][c][0.5]{$0.96$}
 \psfrag{0.94}[c][c][0.5]{$0.94$}
 \psfrag{0.90}[c][c][0.5]{$0.900$}
 \psfrag{0.0}[c][c][0.5]{$0.0$}
 \psfrag{0.2}[c][c][0.5]{$0.2$}
 \psfrag{0.4}[c][c][0.5]{$0.4$}
 \psfrag{0.5}[c][c][0.5]{$0.5$}
 \psfrag{0.6}[c][c][0.5]{$0.6$}
 \psfrag{0.8}[c][c][0.5]{$0.8$}
 \psfrag{1.0}[c][c][0.5]{$1.0$}
 \psfrag{1.5}[c][c][0.5]{$1.5$}
 \psfrag{2.0}[c][c][0.5]{$2.0$}
 \includegraphics[width=12cm,height=10cm]{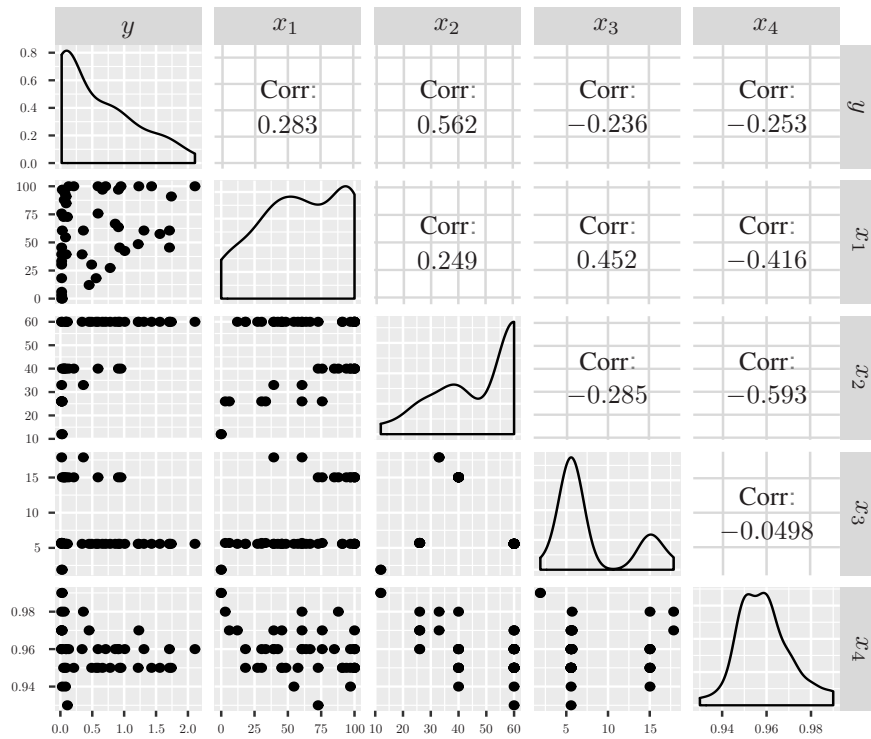}
 \caption{Scatter plot between the explanatory variables and response variable.}
 \label{fig:1}
 \end{figure}

\begin{figure}[H]
 \centering
 \psfrag{0}[c][c][0.3]{0}
 \psfrag{100}[c][c][0.3]{100}
 \psfrag{5}[c][c][0.3]{5}
 \psfrag{10}[c][c][0.3]{10}
 \psfrag{15}[c][c][0.3]{15}
 \psfrag{20}[c][c][0.3]{20}
 \psfrag{25}[c][c][0.3]{25}
 \psfrag{30}[c][c][0.3]{30}
 \psfrag{40}[c][c][0.3]{40}
 \psfrag{50}[c][c][0.3]{50}
  \psfrag{60}[c][c][0.3]{60}
  \psfrag{75}[c][c][0.3]{75}
  \psfrag{80}[c][c][0.3]{80}
  \psfrag{0.06}[c][c][0.3]{0.06}
  \psfrag{0.04}[c][c][0.3]{0.04}
  \psfrag{0.03}[c][c][0.3]{0.03}
  \psfrag{0.02}[c][c][0.3]{0.02}
  \psfrag{0.01}[c][c][0.3]{0.01}
  \psfrag{0.00}[c][c][0.3]{0.00}
  \psfrag{0.900}[c][c][0.3]{0.900}
  \psfrag{0.925}[c][c][0.3]{0.925}
  \psfrag{0.950}[c][c][0.3]{0.950}
  \psfrag{0.975}[c][c][0.3]{0.975}
  \psfrag{density}[c][c][0.5]{Density}
  \psfrag{x1}[c][c][0.5]{Fusarium verticillioides}
  \psfrag{x2}[c][c][0.5]{Time}
  \psfrag{x3}[c][c][0.5]{Precipitation}
  \psfrag{x4}[c][c][0.5]{Water activity}
  \subfigure[\label{fig:y0a}][$x_1$]{\includegraphics[width=3.5cm,height=4cm]{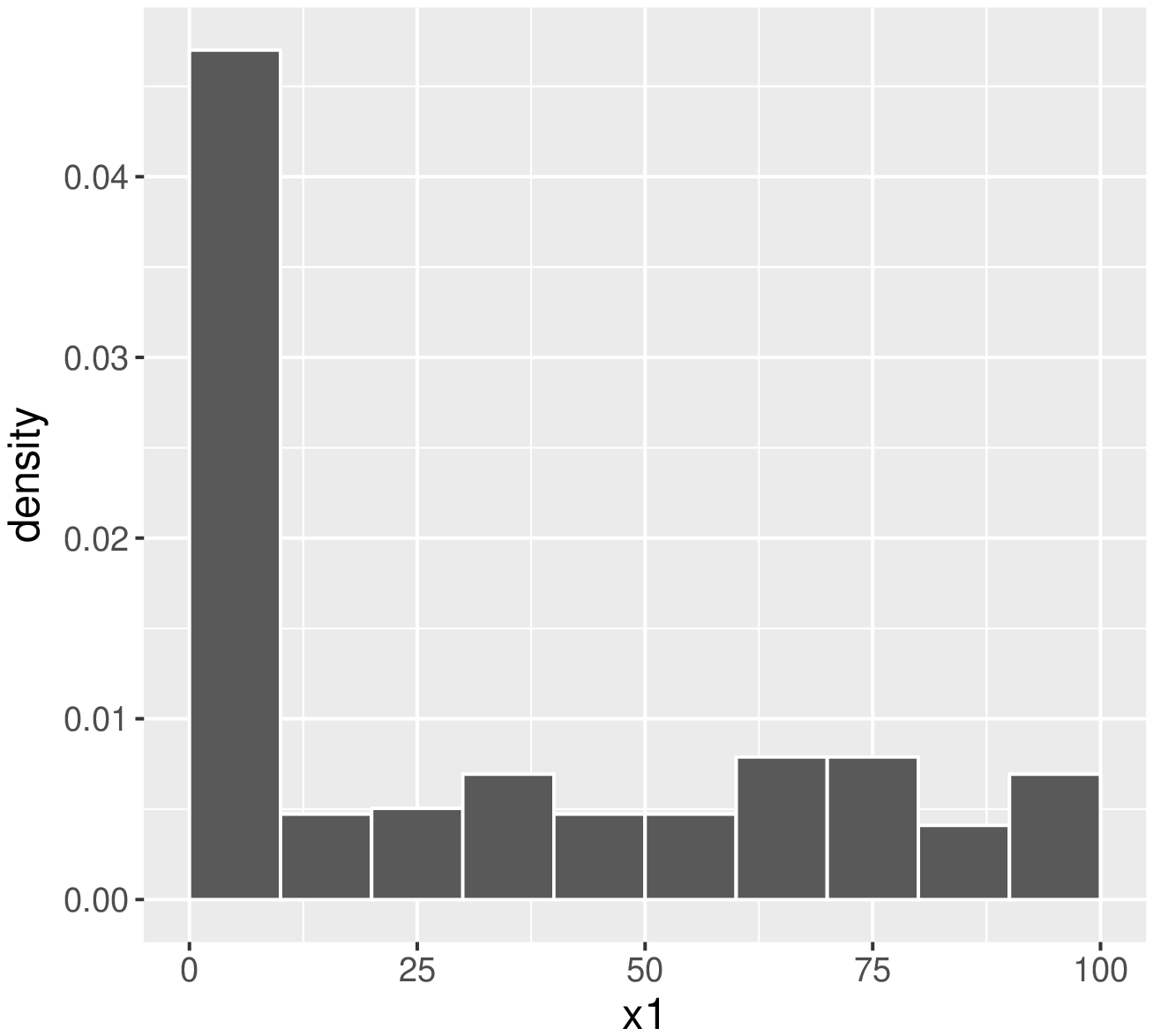}}\qquad
  \subfigure[\label{fig:y0b}][$x_2$]{\includegraphics[width=3.5cm,height=4cm]{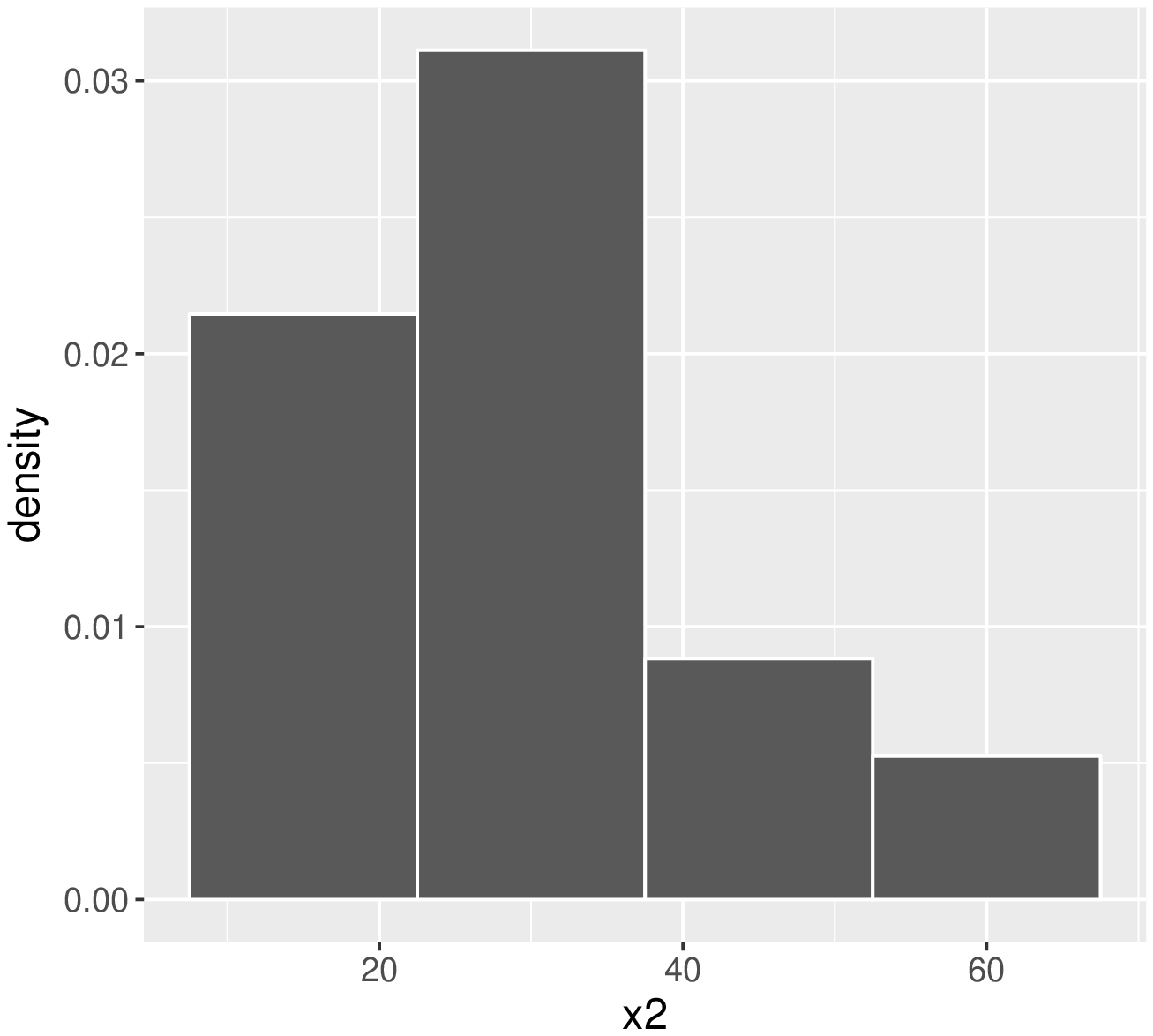}}\qquad
  \subfigure[\label{fig:y0c}][$x_3$]{\includegraphics[width=3.5cm,height=4cm]{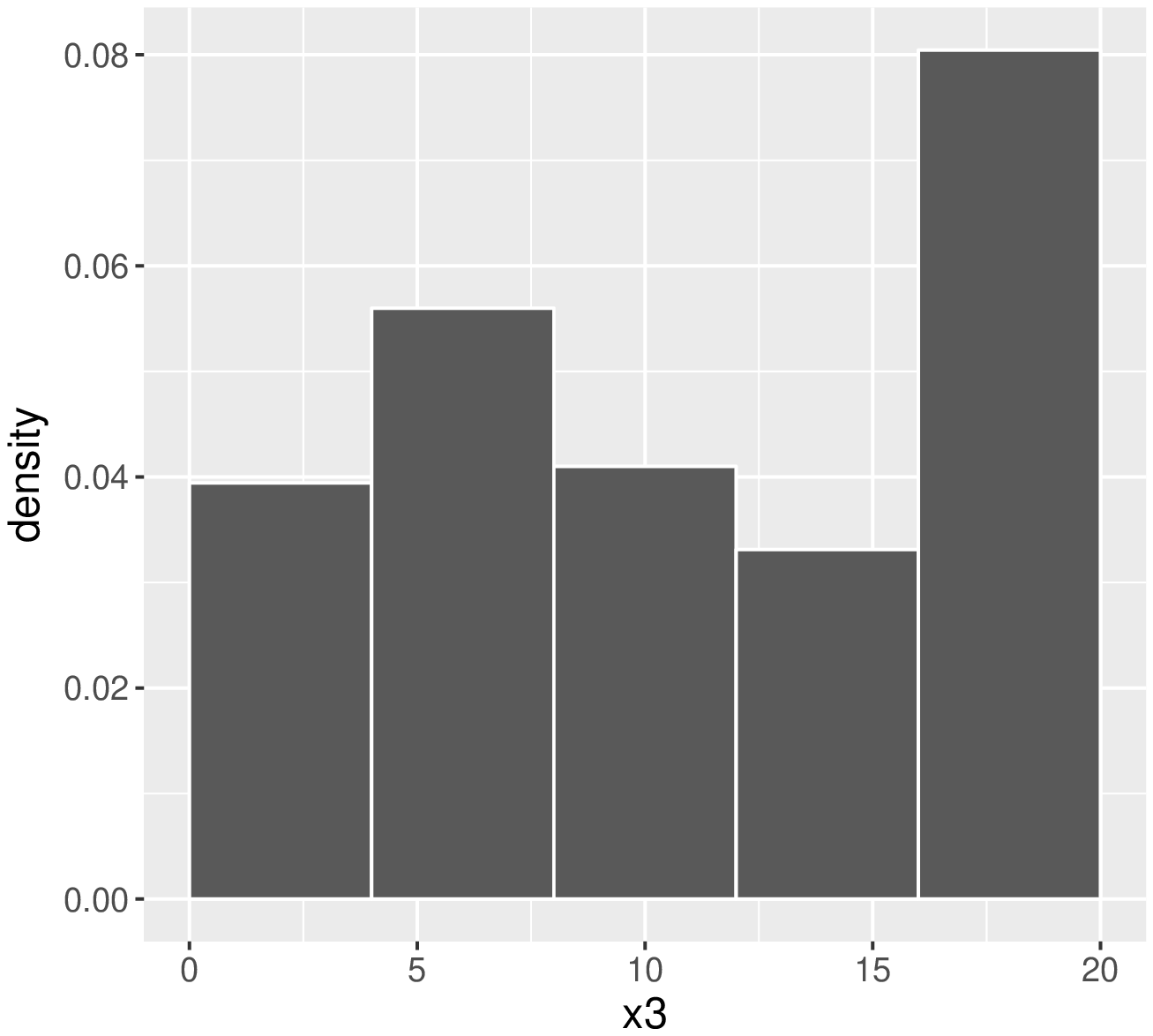}}\qquad
  \subfigure[\label{fig:y0d}][$x_4$]{\includegraphics[width=3.5cm,height=4cm]{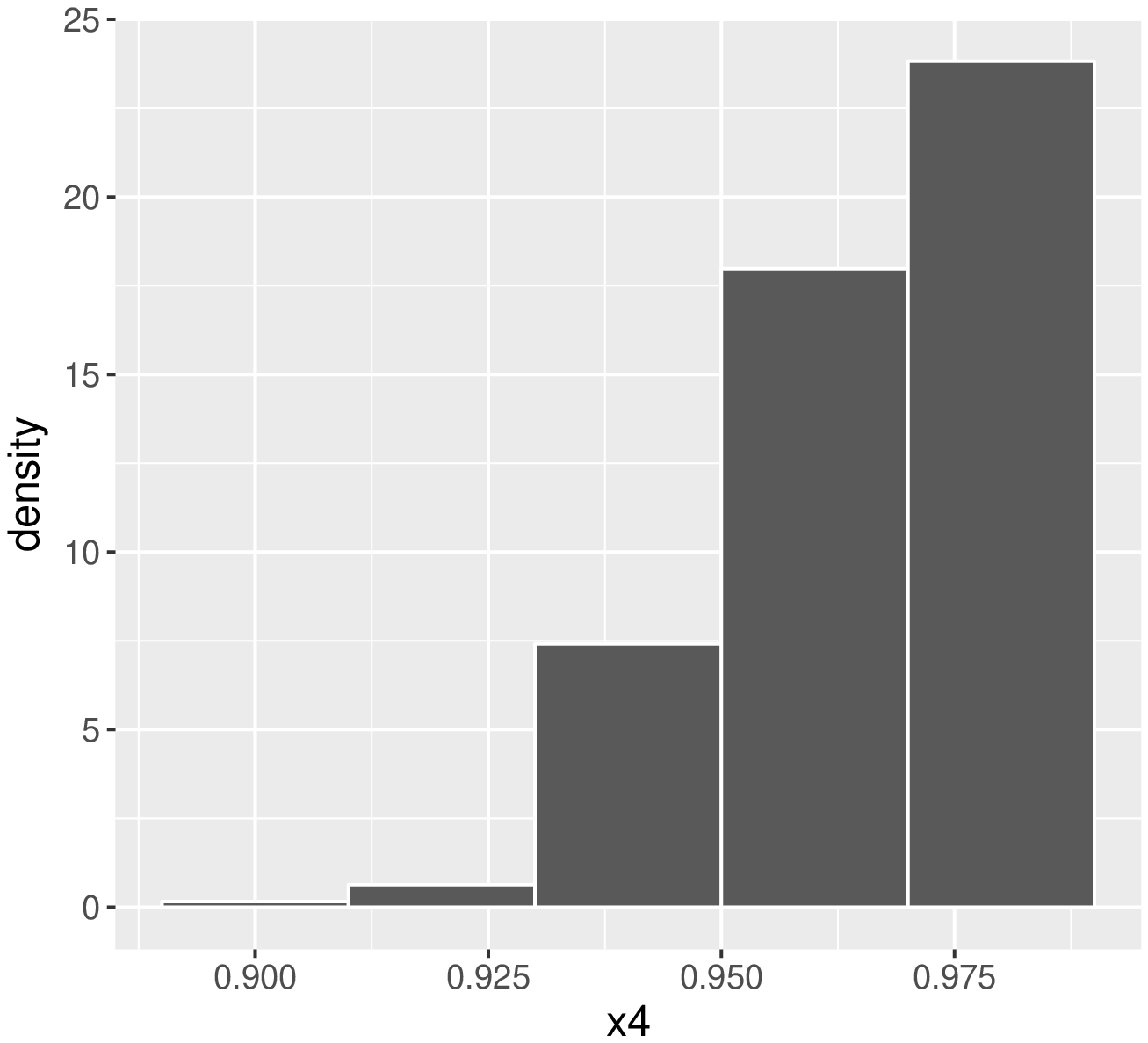}}
  \caption{Zero distributions by explanatory variables.}
  \label{fig:2}
  \end{figure}

Figure~\ref{fig:1} displays the scatterplot between independent and response variables (without zeros).  Figure~\ref{fig:2} shows the zero distribution with relation to the explanatory variables. In addition, we fit different models to see whether there is a need to model the precision of the model and to check whether the initial model should consider the fixed precision. To verify which link functions are more appropriate for fitting the mean and the proportion of zeros, we made several fits and selected the model with the lowest AIC. Finally, we verified that the covariates $x_1$ and $x_2$ are significat for the mean and $x_1$, $x_2$ and $x_3$ are significat for the proportion of zeros.


Based on the initial analysis, we will assume the response $Y_i \sim \textrm{ZABS}(\mu_i,\sigma_i,\nu_i)$ for \textbf{FUMCorn} data. The systematic components of the regression model are expressed as
\begin{eqnarray}\label{eq:modaplic1}
\log(\mu_i) = \beta_0 + \beta_1 x_{i1} +\beta_2 x_{i2}, \quad
 \sigma_i=\alpha_0, \quad
\textrm{probit}(\nu_i) = \gamma_0 + \gamma_1 z_{i1} +  \gamma_2 z_{i2}  +\gamma_3 z_{i3},  \quad i=1,2,\ldots,n, 
\end{eqnarray}
where $\bm \beta=[\beta_0,\beta_1,\beta_2]^\top$ and $\bm \gamma=[\gamma_0,\gamma_1,\gamma_2, \gamma_3]^\top$ being the regression coefficients and $x_{i1} \equiv z_{i1}$, $x_{i2} \equiv z_{i2}$, $x_{i3} \equiv z_{i3}$ and $x_{i4} \equiv z_{i4}$ are the values of the regressor $\mathbf{X}$. We fit the ZABS model 
by using the \textbf{RBS} package presented in Section~\ref{sec:2}. The MLE's of the model parameters, with approximate estimated standard errors, are presented in Table~\ref{tab1:aplic1}.  Note that all the parameters are highly significant.
%
\begin{table}[H] \centering 
  \caption{MLEs, standard error estimates and $p$-values under the ZABS regression model fitted to \textbf{FUMCorn} data.} 
  \label{tab1:aplic1} 
\begin{tabular}{@{\extracolsep{5pt}} ccccc} 
\\[-1.8ex]\hline 
\hline \\[-1.8ex] 
 Parameter& Estimate & Std. Error & t value & Pr(\textgreater \textbar t\textbar ) \\ 
\hline \\[-1.8ex] 
$\beta_0$ & \multicolumn{1}{r}{$$-$4.726$} & $0.692$ & $$-$6.830$ & $<0.01$ \\ 
$\beta_1$& \multicolumn{1}{r}{$0.015$} & $0.005$ & $2.816$ & $<0.01$ \\ 
$\beta_2$ & \multicolumn{1}{r}{$0.057$} & $0.012$ & $4.967$ & $<0.01$ \\ 
$\gamma_0$ & \multicolumn{1}{r}{$1.934$} & $0.331$ & $5.850$ & $<0.01$ \\ 
$\gamma_1$ & \multicolumn{1}{r}{$-0.014 $} & $0.004$ & $-3.560$ & $<0.01$ \\ 
$\gamma_2$ & \multicolumn{1}{r}{$-0.030$} & $0.007$ & $-4.173$ & $<0.01$ \\ 
$\gamma_3$ & \multicolumn{1}{r}{$0.096$} & $0.024$ & $4.079$ & $<0.01$ \\ 
$\alpha_0$ & \multicolumn{1}{r}{$1.127$} & $0.233$ & $\times$ & $\times$ \\ 
\hline \\[-1.8ex] 
\end{tabular} 
\end{table} 

\begin{figure}[H]
  \centering
  \psfrag{td}[c][c]{\tiny Empirical Quantile}
  \psfrag{r}[c][c]{\tiny Theorical Quantile}
  \psfrag{fv}[c][c][0.8]{\tiny Fitted Values}
  \psfrag{res}[c][c][0.8]{\tiny Quantile Residuals}
  \psfrag{x1}[c][c][0.8]{\tiny $x_1$}
  \psfrag{x2}[c][c][0.8]{\tiny $x_2$}
  \psfrag{0}[c][c][0.8]{\tiny 0}
  \psfrag{0.00}[c][c][0.8]{\tiny 0.00}
  \psfrag{0.25}[c][c][0.8]{\tiny 0.25}
  \psfrag{0.50}[c][c][0.8]{\tiny 0.50}
  \psfrag{0.75}[c][c][0.8]{\tiny 0.75}
  \psfrag{1.00}[c][c][0.8]{\tiny 1.00}
  \psfrag{1.25}[c][c][0.8]{\tiny 1.25}
  \psfrag{1.2}[c][c][0.8]{\tiny 1.2}
 \psfrag{-3}[c][c][0.8]{\tiny -3}
  \psfrag{-4}[c][c][0.8]{\tiny -4}
  \psfrag{4}[c][c][0.8]{\tiny 4}
  \psfrag{-2}[c][c][0.8]{\tiny -2}
  \psfrag{-1}[c][c][0.8]{\tiny -1}
  \psfrag{1}[c][c][0.8]{\tiny 1}
  \psfrag{2}[c][c][0.8]{\tiny 2}
  \psfrag{3}[c][c][0.8]{\tiny 3}
  \psfrag{10}[c][c][0.8]{\tiny 10}
  \psfrag{20}[c][c][0.8]{\tiny 20}
  \psfrag{25}[c][c][0.8]{\tiny 25}
  \psfrag{30}[c][c][0.8]{\tiny 30}
  \psfrag{40}[c][c][0.8]{\tiny 40}
  \psfrag{50}[c][c][0.8]{\tiny 50}
  \psfrag{60}[c][c][0.8]{\tiny 60}
  \psfrag{75}[c][c][0.8]{\tiny 75}
  \psfrag{80}[c][c][0.8]{\tiny 80}
  \psfrag{100}[c][c][0.8]{\tiny 100}
  \subfigure[][\label{fig:3a}]{\includegraphics[width=3.5cm,height=4cm]{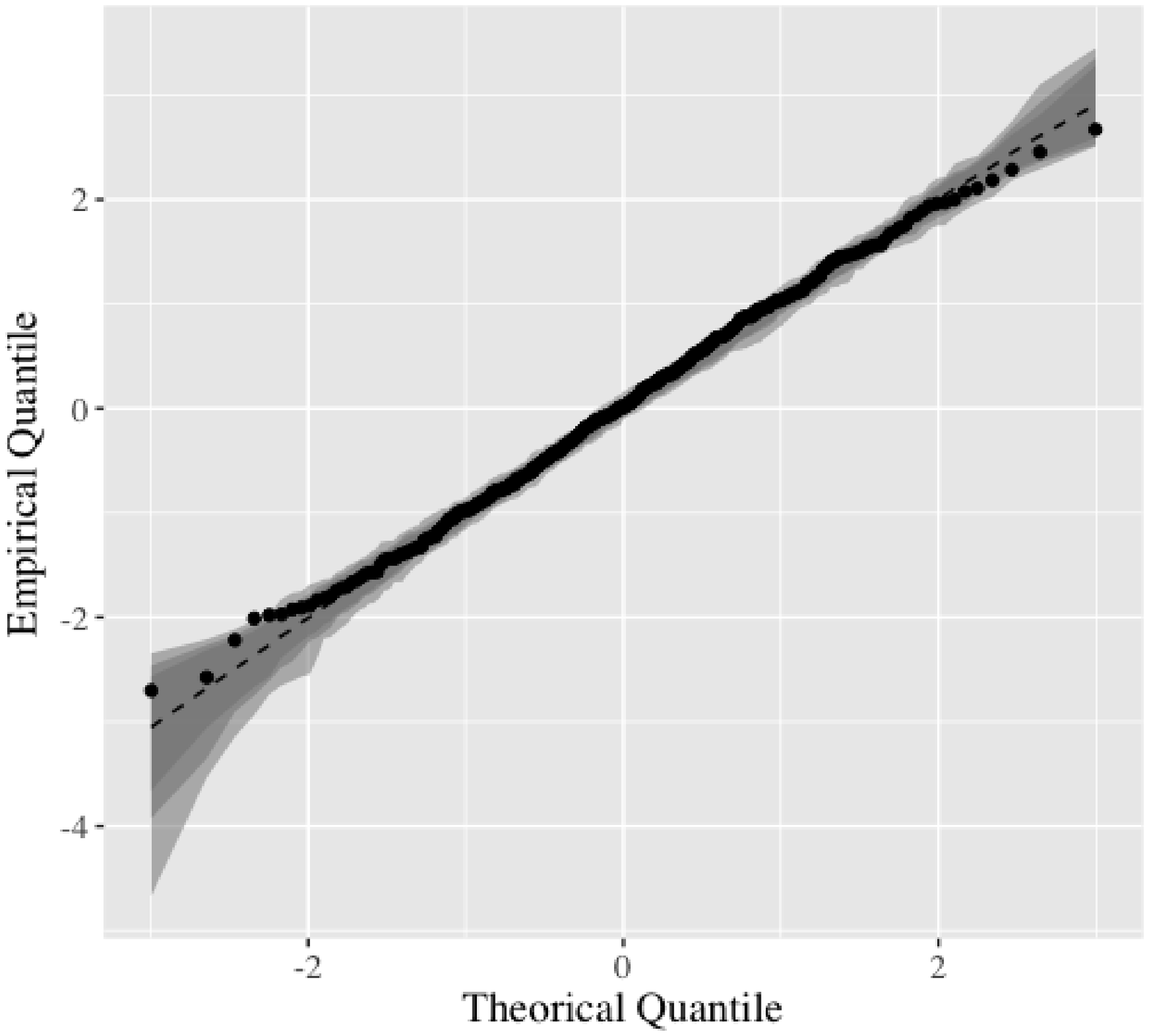}}\qquad
  \subfigure[][\label{fig:3b}]{\includegraphics[width=3.5cm,height=4cm]{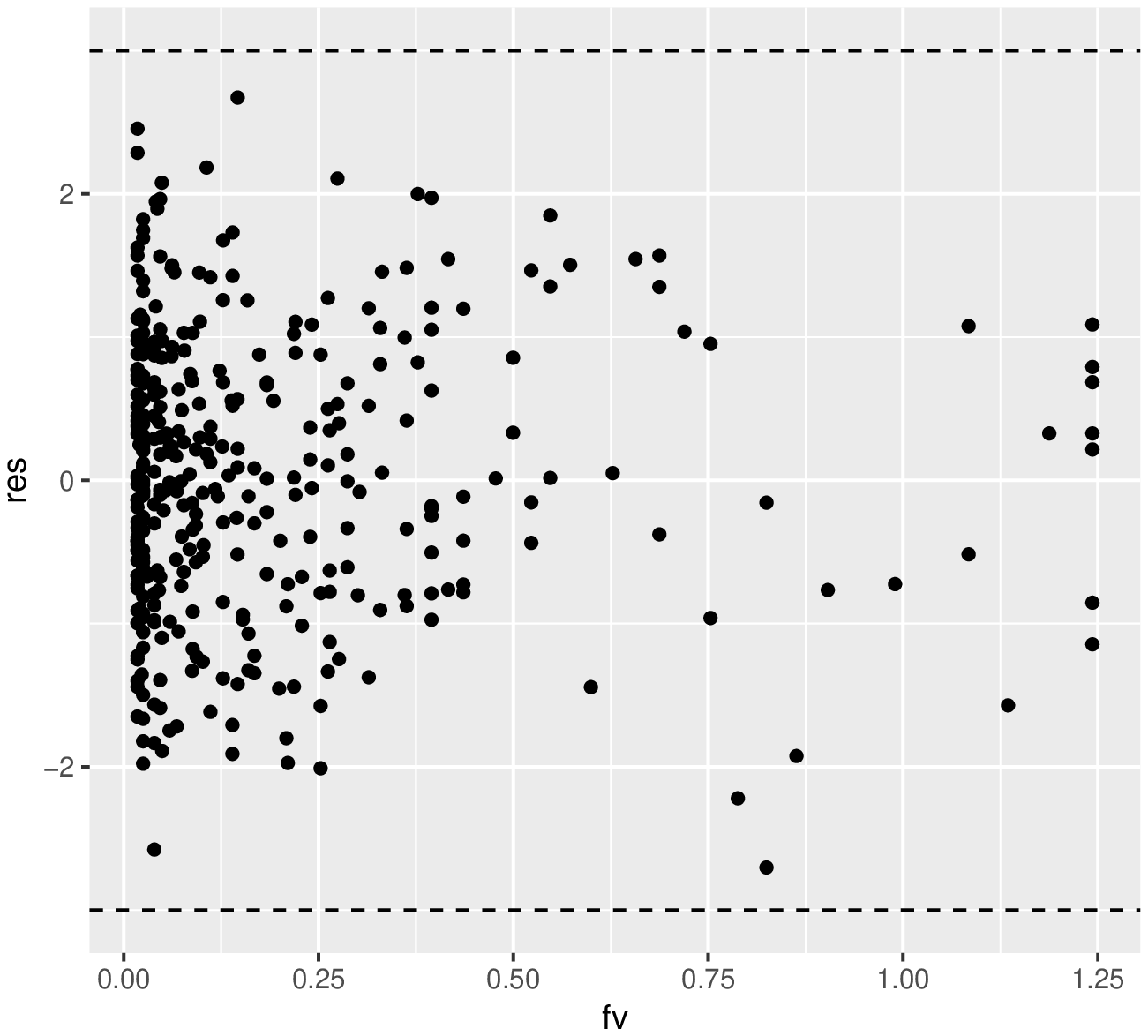}}\qquad
  \subfigure[][\label{fig:3c}]{\includegraphics[width=3.5cm,height=4cm]{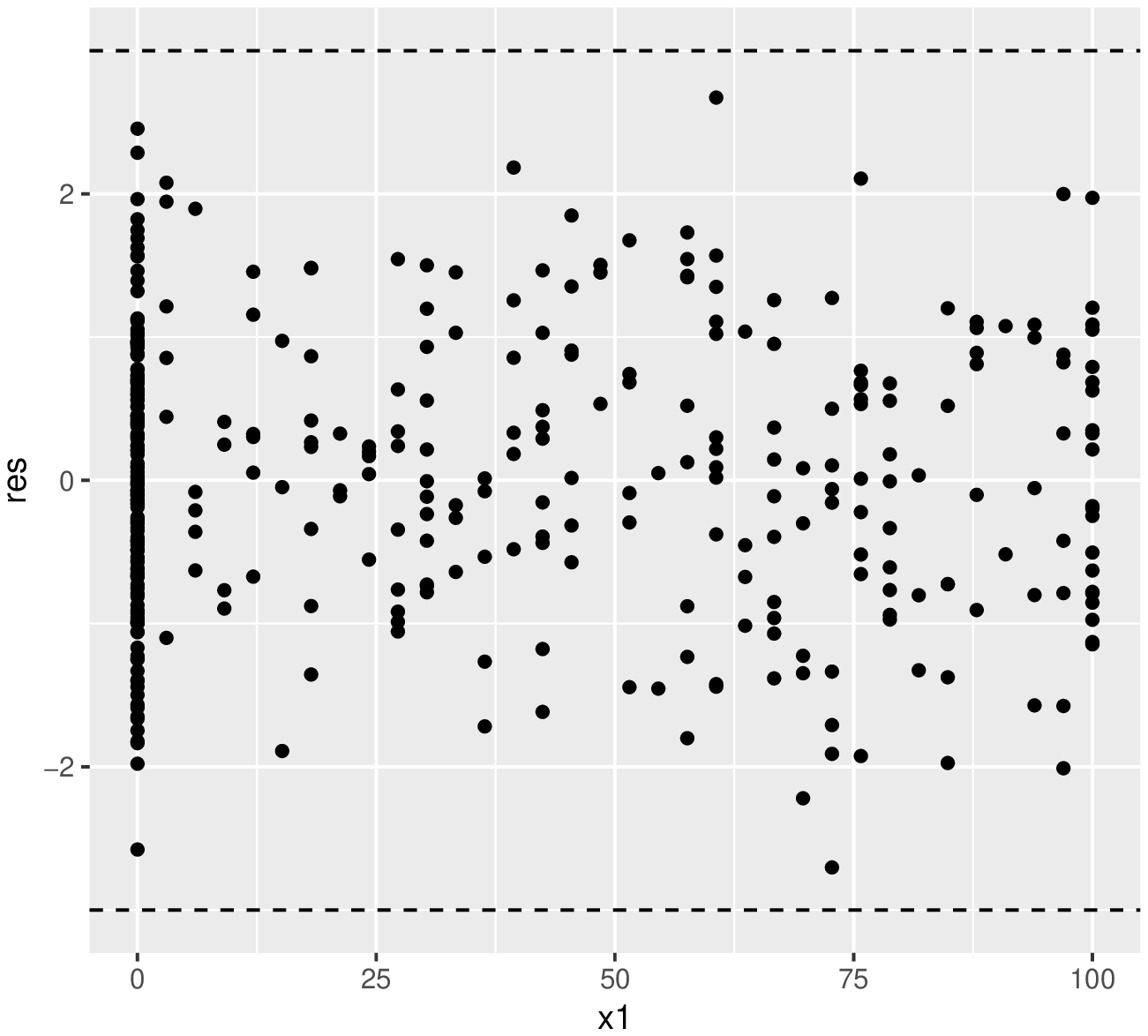}}\qquad
  \subfigure[][\label{fig:3d}]{\includegraphics[width=3.5cm,height=4cm]{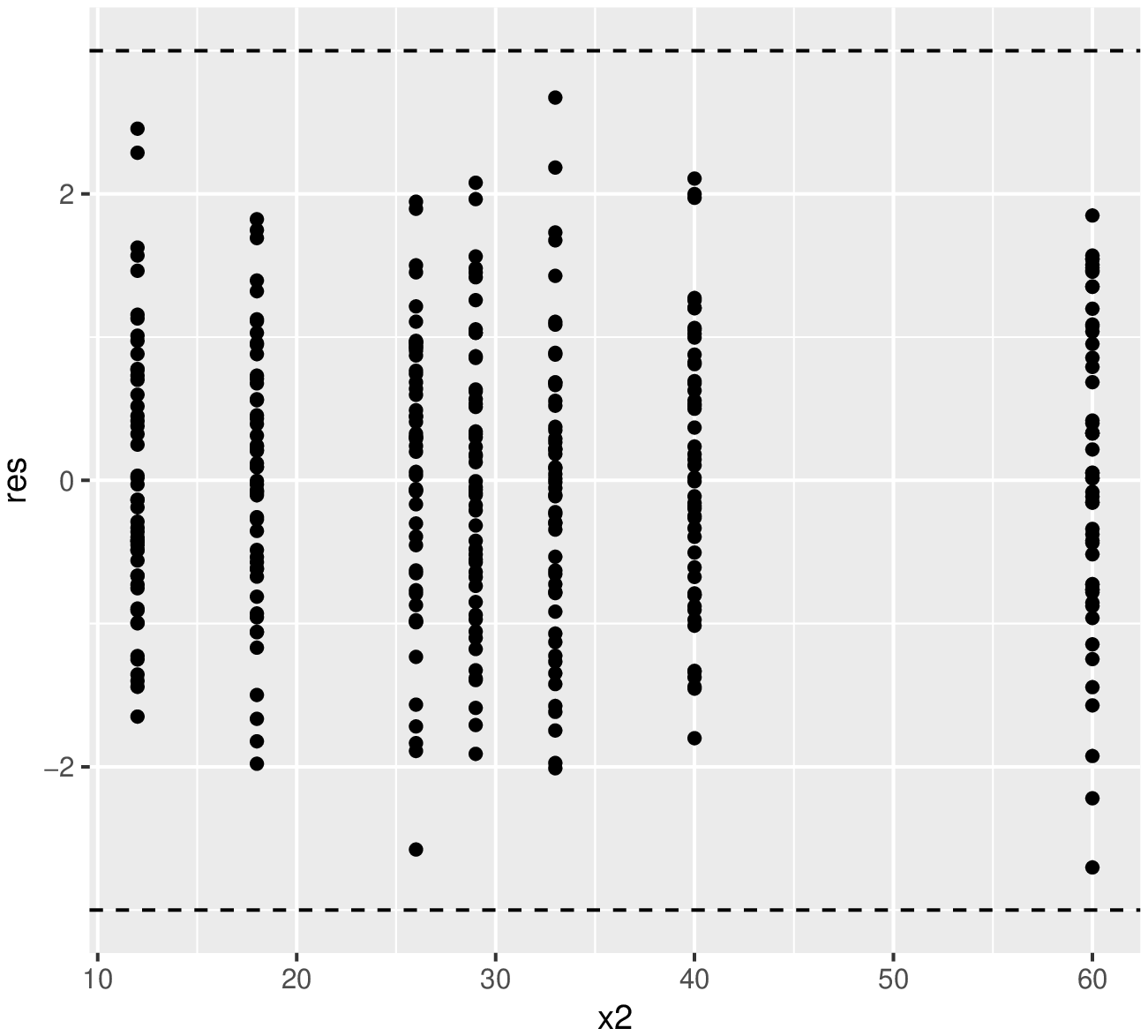}}
  \caption{QQ-plot with simulated envelope of quantile residuals (a), fitted values versus quantile residuals (b), $x_1$ versus quantile residuals (c) and $x_2$ versus quantile residuals (d).}
  \label{fig:3}
  \end{figure}

 The assumptions of the model are verified by residual analysis based on \textbf{FUMCorn} data. 
 The normal probability plot with envelope for the quantile residual, showed in Figure~\ref{fig:3a},
 is used to verify the distributional assumption of the model. This Figure does not show unusual 
 features, so that the assumption that the response variable follows a ZABS distribution 
 does not seem to be unsuitable.  Figures~\ref{fig:3a}-\ref{fig:3d} suggest the model is well fitted and outliers are not detected.

\begin{figure}[H]
  \begin{center}
  \psfrag{Cibeta}[c][c][0.8]{\tiny $\mathrm{C}_i$}
  \psfrag{Cialpha}[c][c][0.8]{\tiny $\mathrm{C}_i$}
  \psfrag{Cigamma}[c][c][0.8]{\tiny $\mathrm{C}_i$}
  \psfrag{Citheta}[c][c][0.8]{\tiny $\mathrm{C}_i(\theta)$}
  \psfrag{I}[c][c][0.8]{\tiny index}
  \psfrag{1.00}[c][c][0.8][90]{\tiny 1.00}
  \psfrag{0.75}[c][c][0.8][90]{\tiny 0.75}
  \psfrag{0.50}[c][c][0.8][90]{\tiny 0.50}
  \psfrag{0.25}[c][c][0.8][90]{\tiny 0.25}
  \psfrag{0.00}[c][c][0.8][90]{\tiny 0.00}
  \psfrag{0}[c][c][0.8]{\tiny 0}
  \psfrag{228}[c][c]{\tiny 228}
  \psfrag{7}[r][l]{\tiny 7}
  \psfrag{100}[c][c][0.8]{\tiny 100}
  \psfrag{200}[c][c][0.8]{\tiny 200}
  \psfrag{300}[c][c][0.8]{\tiny 300}
  \subfigure[\label{fig:cia}][$\bm \beta$]{\includegraphics[width=3.5cm,height=4cm]{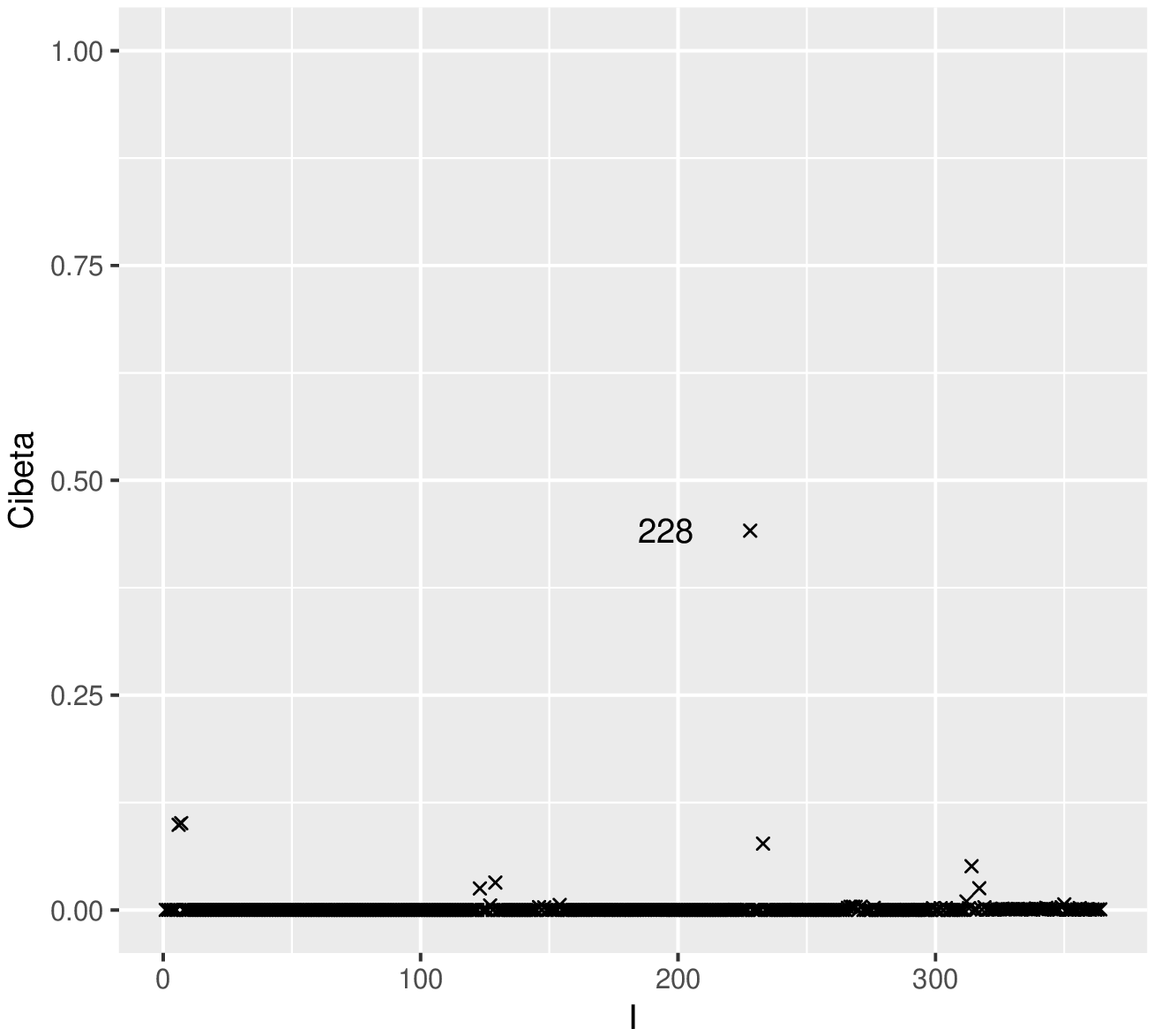}}\qquad
  \subfigure[\label{fig:cib}][$\bm \alpha$]{\includegraphics[width=3.5cm,height=4cm]{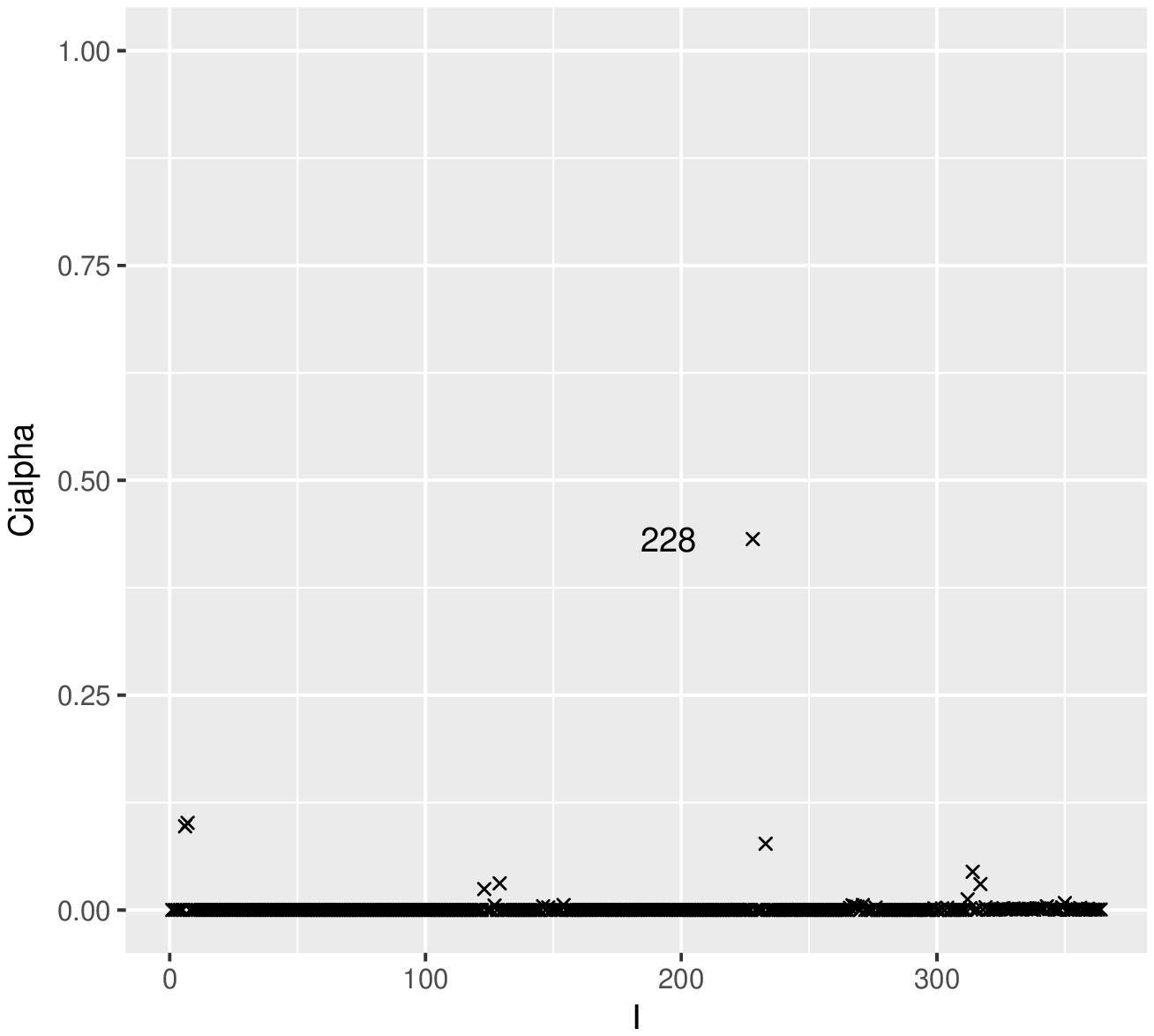}}\qquad
  \subfigure[\label{fig:cic}][$\bm \gamma$]{\includegraphics[width=3.5cm,height=4cm]{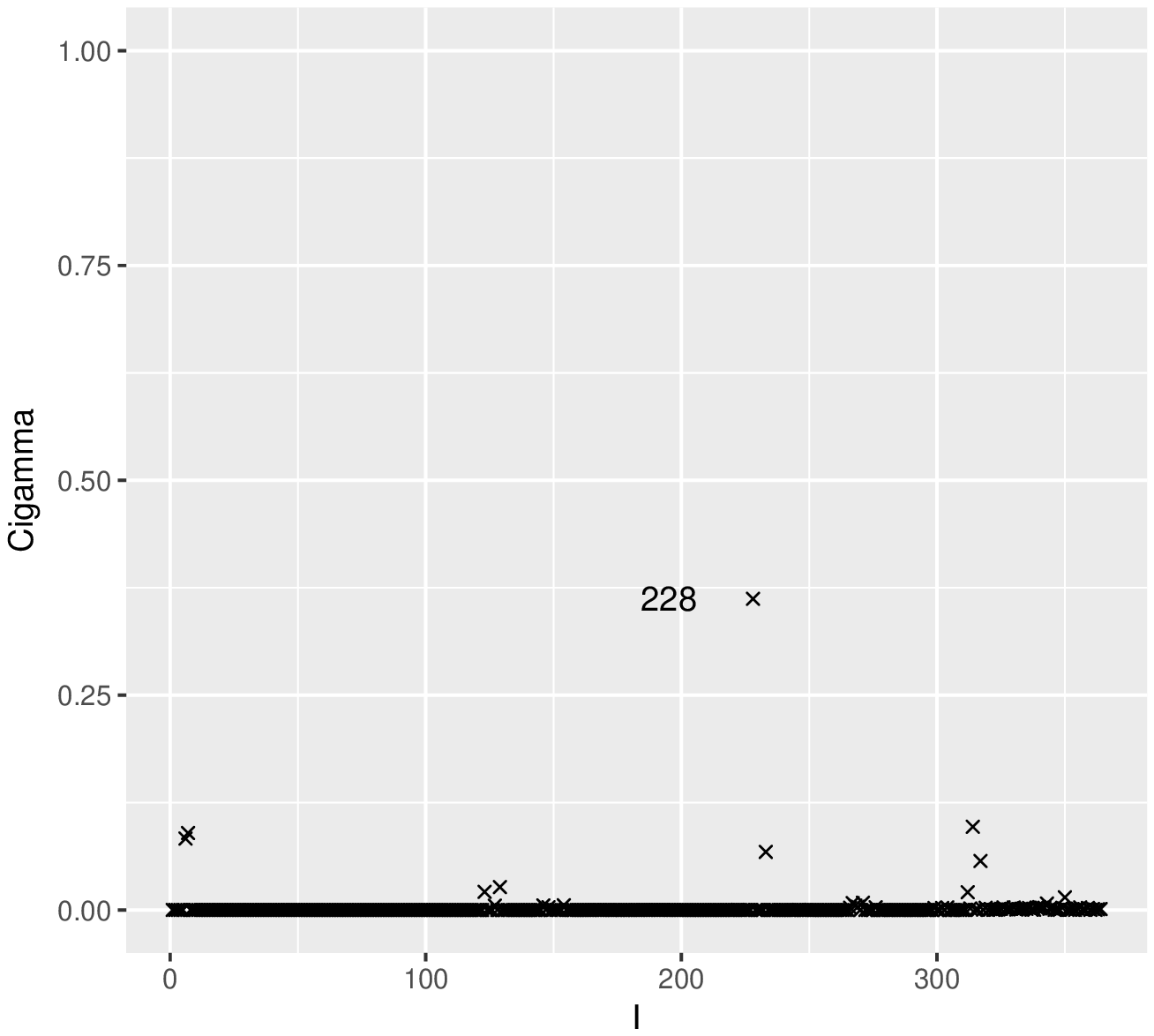}}\qquad
  \subfigure[\label{fig:cid}][$\bm \theta$]{\includegraphics[width=3.5cm,height=4cm]{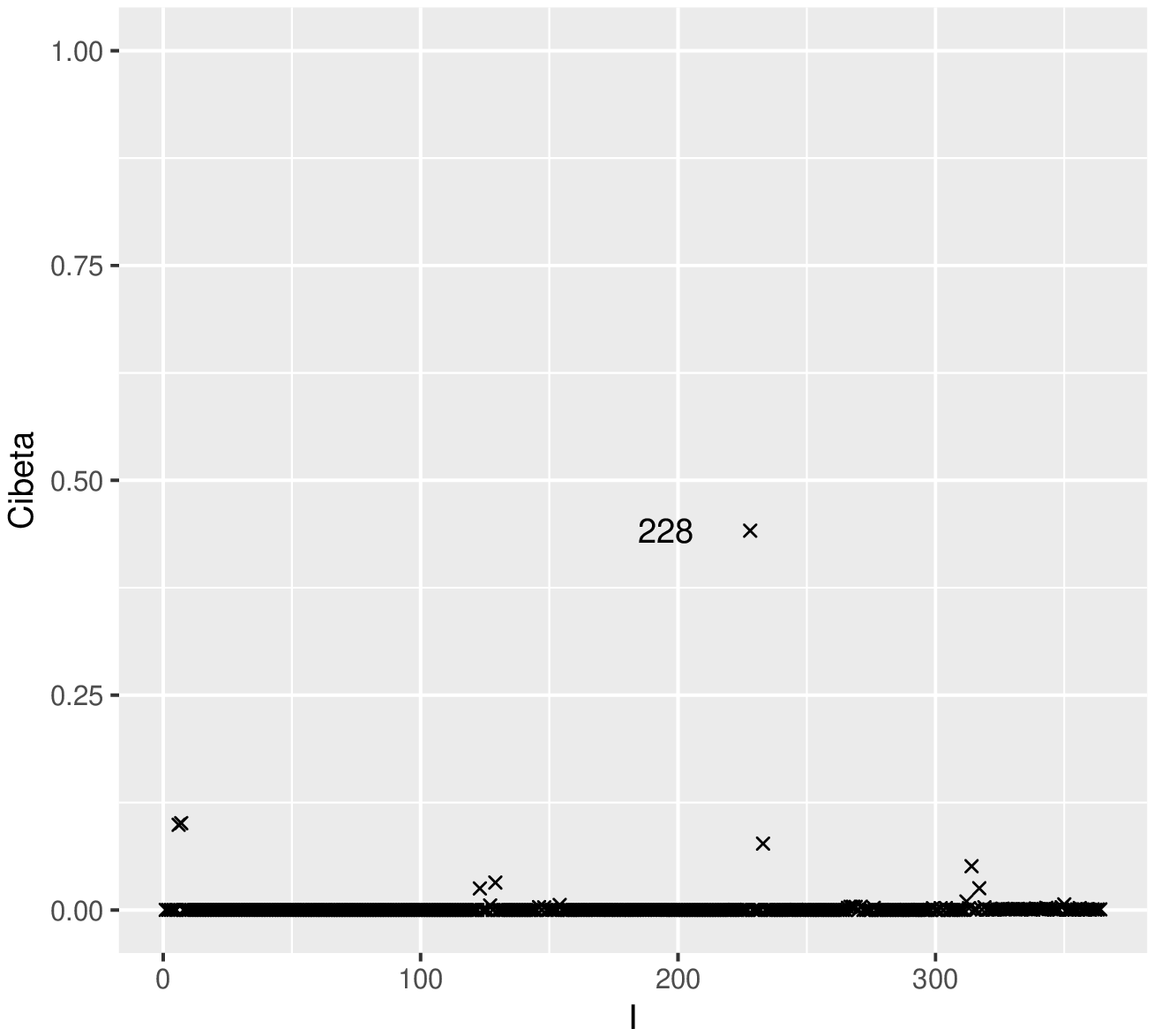}}
  \caption{$\mathrm{C}_i$'s plots.}
  \label{fig:ci}
  \end{center}
  \end{figure}

\begin{figure}[H]
  \begin{center}
  \psfrag{dmbeta}[c][c][0.8]{\tiny $|\mathbf{d}_\text{max}|$}
  \psfrag{dmalpha}[c][c][0.8]{\tiny $|\mathbf{d}_\text{max}|$}
  \psfrag{dmgamma}[c][c][0.8]{\tiny $|\mathbf{d}_\text{max}|$}
  \psfrag{dmtheta}[c][c][0.8]{\tiny $|\mathbf{d}_\text{max}|$}
  \psfrag{I}[c][c][0.8]{\tiny index}
  \psfrag{1.00}[c][c][0.8][90]{\tiny 1.00}
  \psfrag{0.75}[c][c][0.8][90]{\tiny 0.75}
  \psfrag{0.50}[c][c][0.8][90]{\tiny 0.50}
  \psfrag{0.25}[c][c][0.8][90]{\tiny 0.25}
  \psfrag{0.00}[c][c][0.8][90]{\tiny 0.00}
  \psfrag{0}[c][c][0.8]{\tiny 0}
  \psfrag{2}[l][l]{\tiny 2}
  \psfrag{20}[c][r]{\tiny 20}
  \psfrag{100}[c][c][0.8]{\tiny 100}
  \psfrag{200}[c][c][0.8]{\tiny 200}
  \psfrag{300}[c][c][0.8]{\tiny 300}
  \subfigure[\label{fig:lma}][$\bm \beta$]{\includegraphics[width=3.5cm,height=4cm]{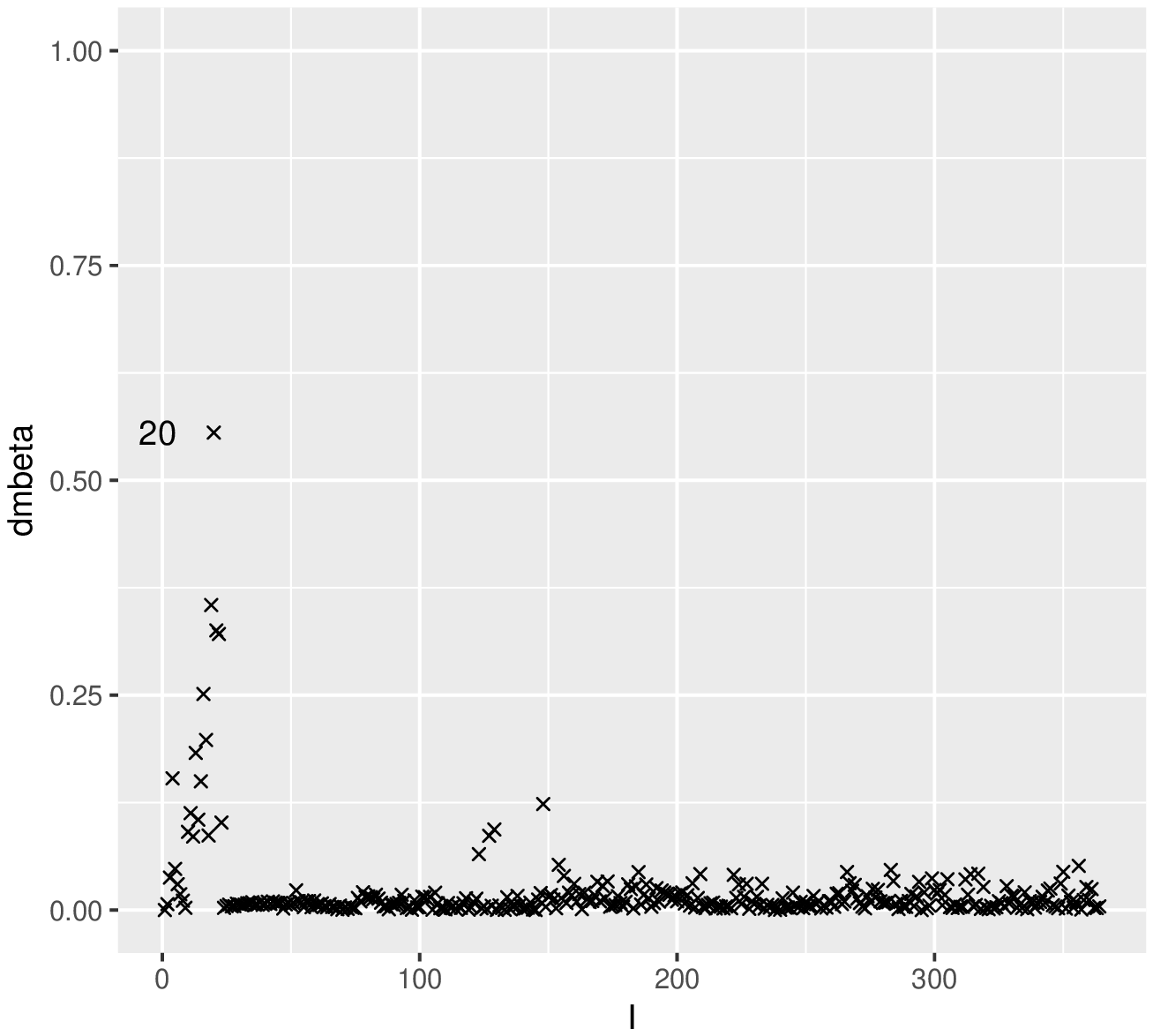}}\qquad
  \subfigure[\label{fig:lmb}][$\bm \alpha$]{\includegraphics[width=3.5cm,height=4cm]{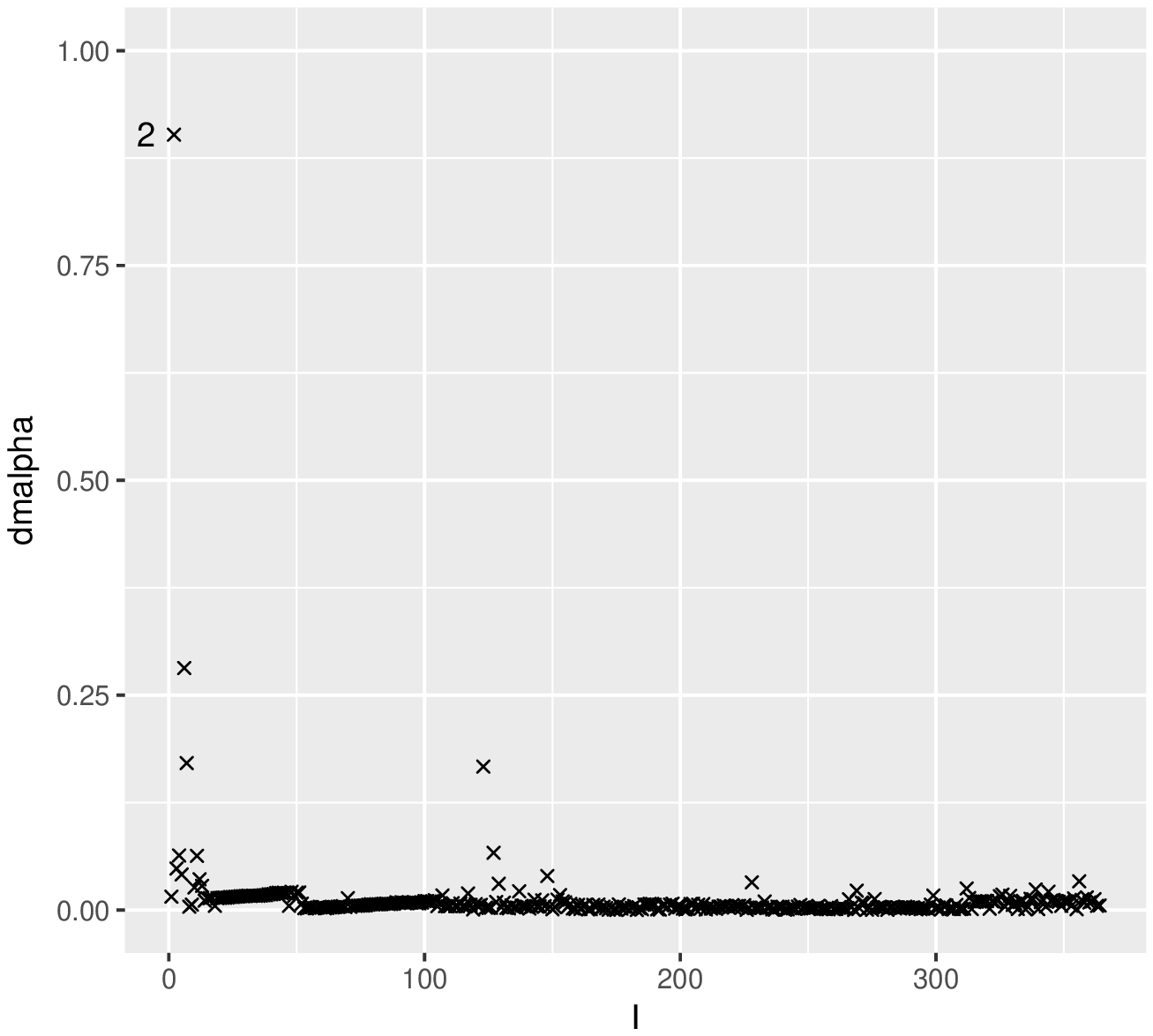}}\qquad
  \subfigure[\label{fig:lmc}][$\bm \gamma$]{\includegraphics[width=3.5cm,height=4cm]{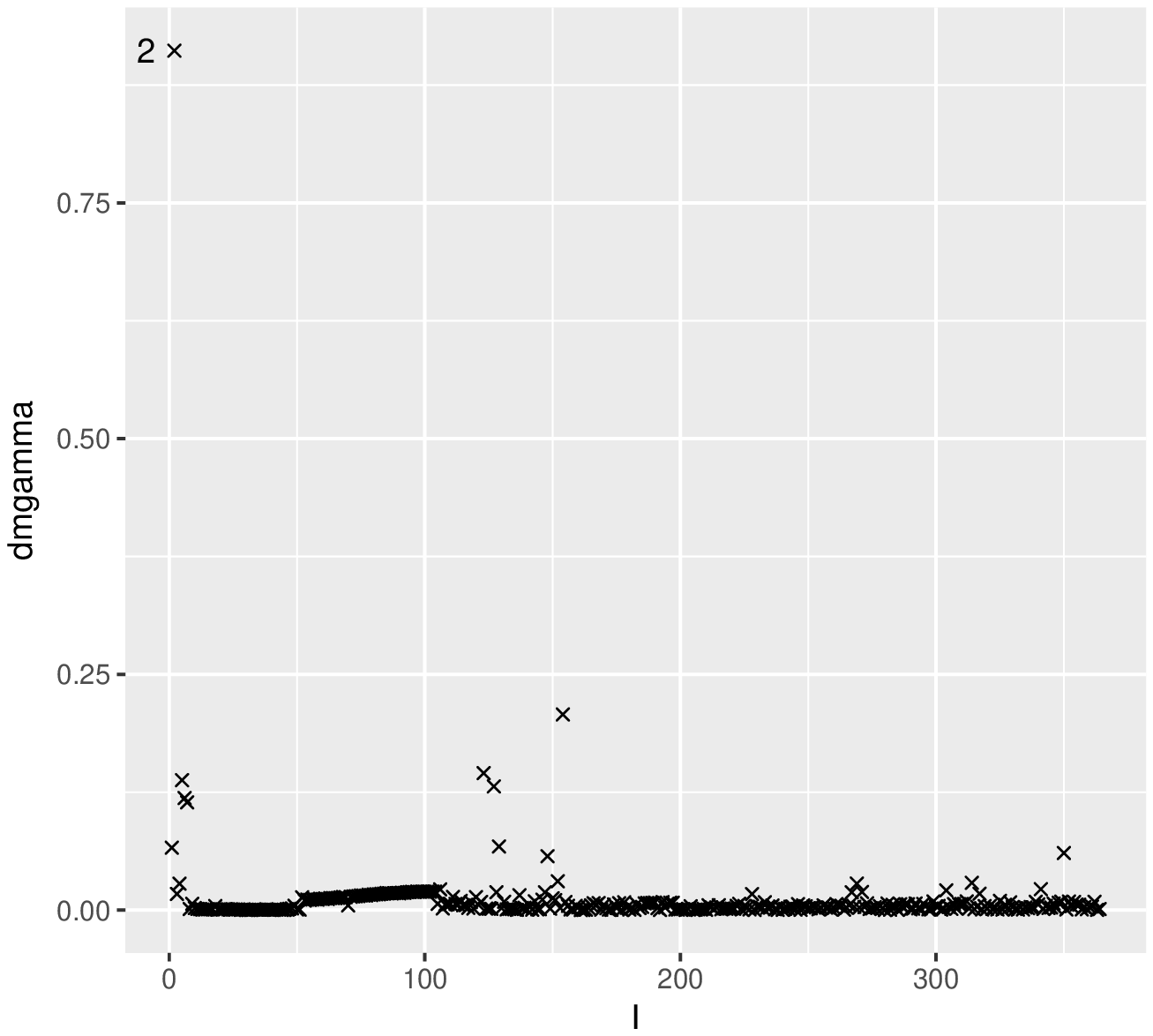}}\qquad
  \subfigure[\label{fig:lmd}][$\bm \theta$]{\includegraphics[width=3.5cm,height=4cm]{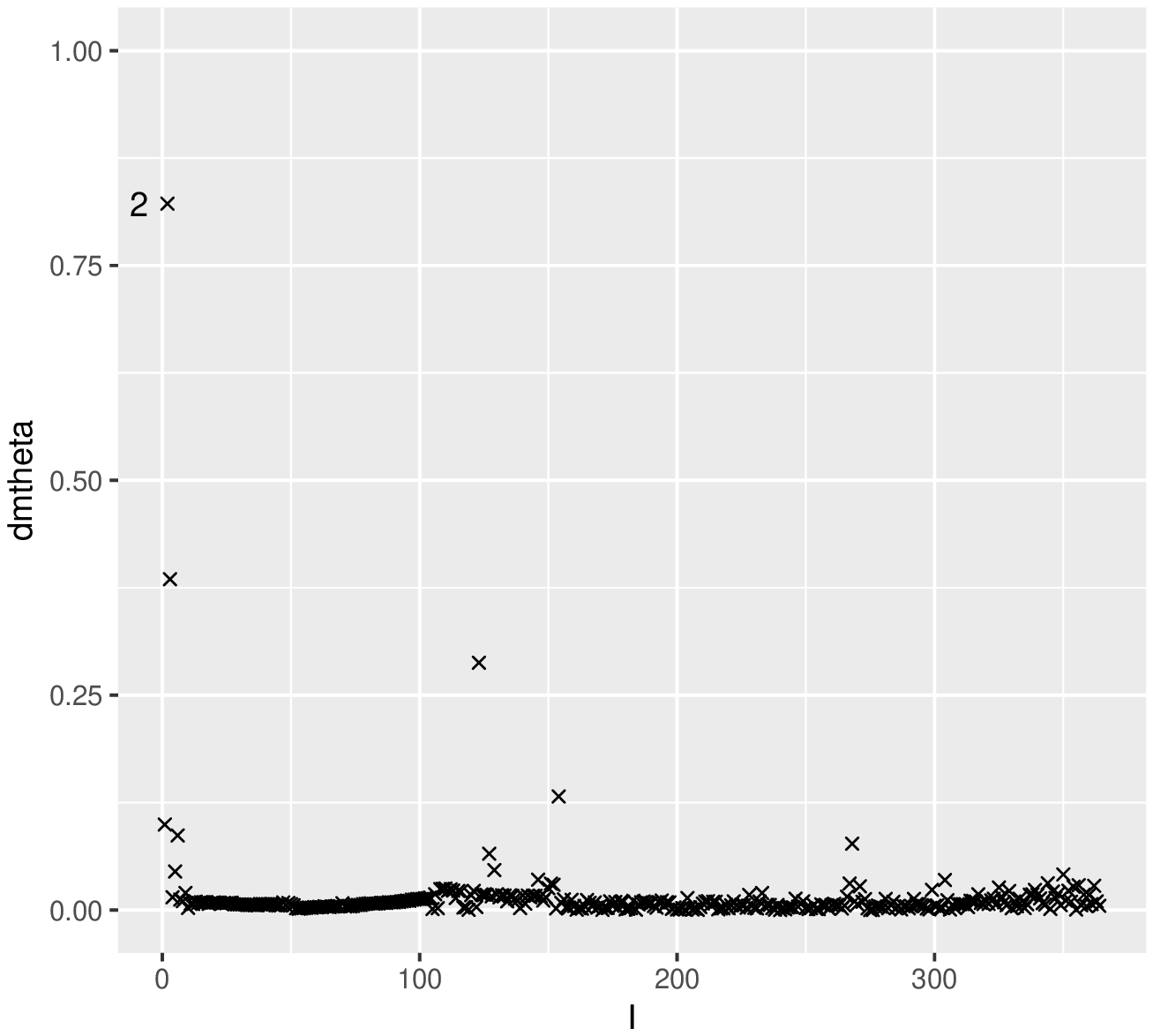}}
  \caption{Index plots of $|\mathbf{d}_\text{max}|$.}
  \label{fig:lm}
  \end{center}
 \end{figure}

 Influence diagnostics for the ZABS regression model are presented in Figure \ref{fig:ci}. Figure \ref{fig:ci} displays index plots of $\mathrm{C}_i$, from where observation \#228 was detected as potentially influential. Note in Figure~\ref{fig:lm} that in addition to the observation \#228, observations \#2 and \#20 appear as potentially influential.
We now investigate the impact on the model inference when the cases detected as potentially influential in the diagnostic analysis are removed. Then, we again estimate the model parameters after removing the sets of observations $S_1= \{2\}$, $S_2=\{20\}$, $S_3=\{228\}$, $S_4=\{2,20\}$, $S_5=\{2,228\}$, $S_6=\{20,228\}$ and $S_7=\{2,20,228\}$. We eliminated those most influential observations and refitted the models. Finally, inferential changes are not detected when the observations (set of observations) are removed.
%

\subsection{BS nonlinear regression model}
\begin{figure}[ht]
  \begin{center}
  \psfrag{N}[c][c][0.8]{\tiny Number of cycles to failure ($\times 100$)}
  \psfrag{W}[c][c][0.8]{\tiny Work per cycle ($mJ/m^3$)}
  \psfrag{Residuals}[c][c]{\tiny Empirical quantile}
  \psfrag{perc}[c][c]{\tiny Theoretical quantile}
  \psfrag{0}[c][c]{\tiny 0}
  \psfrag{-3}[c][c]{\tiny -3}
  \psfrag{-2}[c][c]{\tiny -2}
  \psfrag{-1}[c][c]{\tiny -1}
  \psfrag{1}[c][c]{\tiny 1}
  \psfrag{2}[c][c]{\tiny 2}
  \psfrag{3}[c][c]{\tiny 3}
  \psfrag{10}[c][c]{\tiny 10}
  \psfrag{20}[c][c]{\tiny 20}
  \psfrag{25}[c][c]{\tiny 25}
  \psfrag{30}[c][c]{\tiny 30}
  \psfrag{40}[c][c]{\tiny 40}
  \psfrag{50}[c][c]{\tiny 50}
  \psfrag{60}[c][c]{\tiny 60}
  \psfrag{75}[c][c]{\tiny 75}
  \psfrag{80}[c][c]{\tiny 80}
  \psfrag{100}[c][c]{\tiny 100}
  \subfigure[][\label{fig:7a}]{\includegraphics[width=3.5cm,height=4cm]{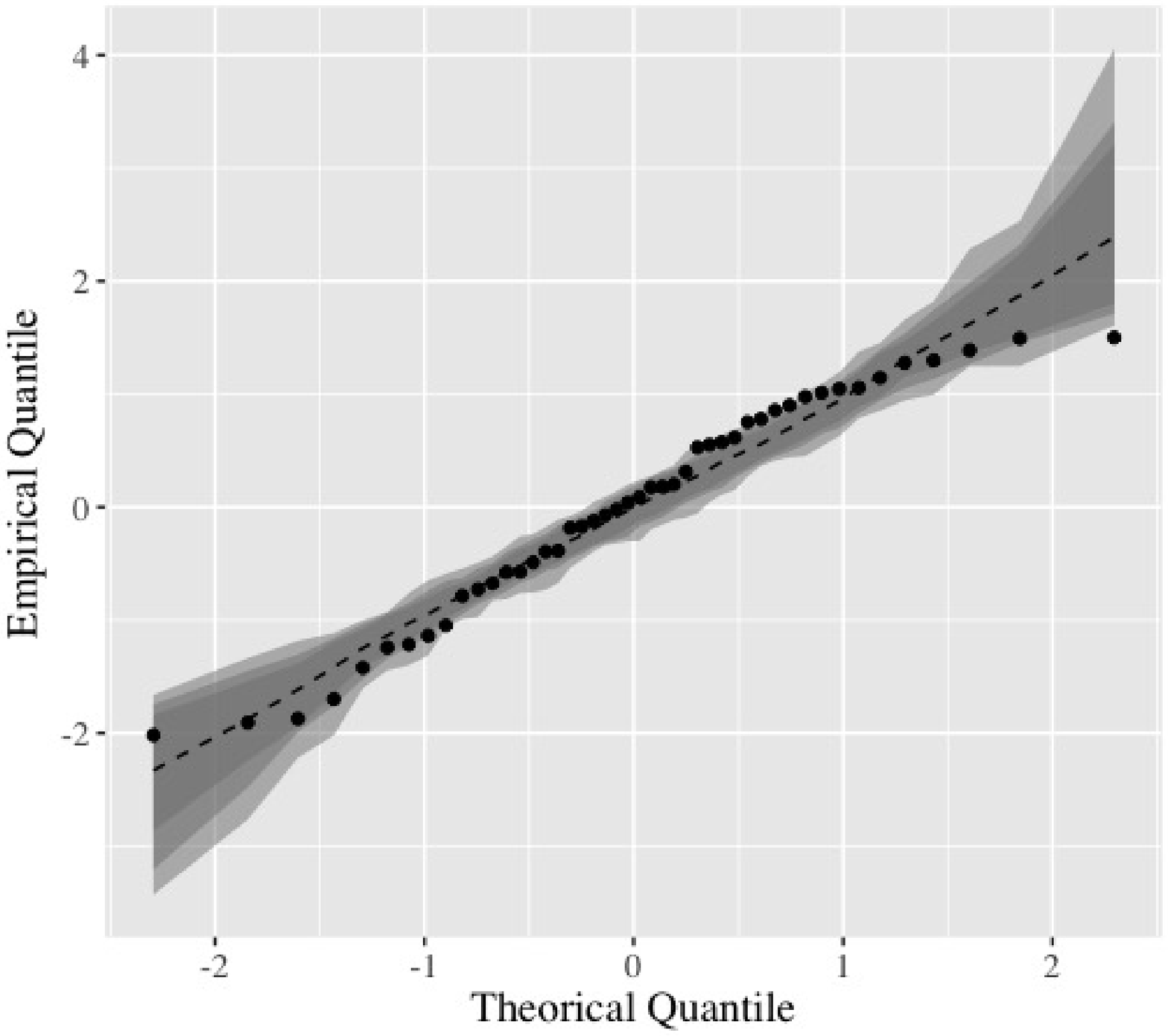}}\qquad
  \subfigure[][\label{fig:7b}]{\includegraphics[width=3.5cm,height=4cm]{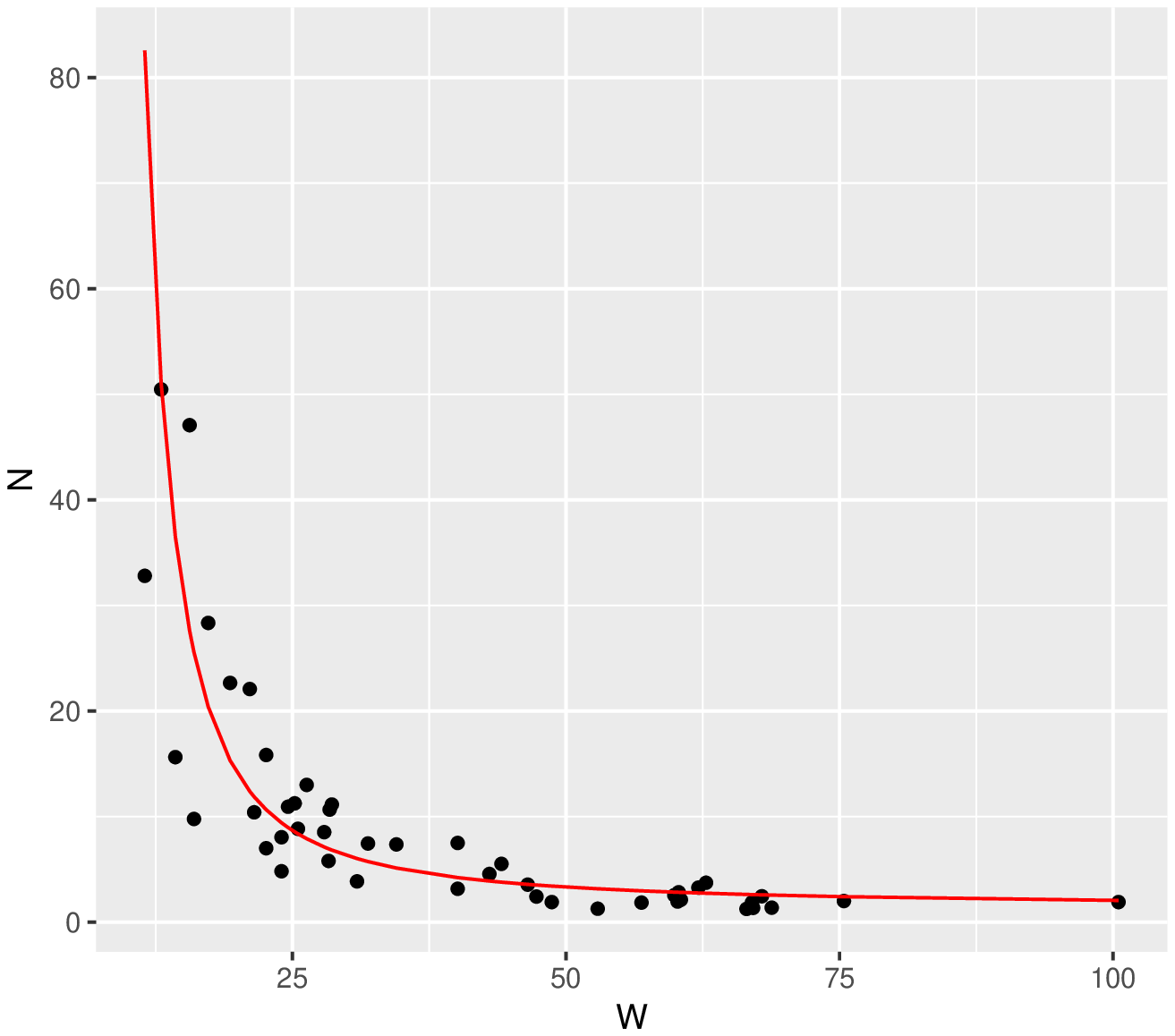}}
  \caption{QQ-plot with simulated envelope of quantile residuals and scatter-plot with fitted model.}
  \label{fig:7}
  \end{center}
  \end{figure}

\begin{figure}[ht]
  \begin{center}
  \psfrag{N}[c][c]{\tiny Number of cycles to failure ($\times 100$)}
  \psfrag{ind}[c][c]{\tiny Index}
  \psfrag{res}[c][c]{\tiny Quantile Residuals}
  \psfrag{fv}[c][c]{\tiny Fitted Values}
  \psfrag{0}[c][c][1][90]{\tiny 0}
  \psfrag{-2}[c][c][1][90]{\tiny -2}
  \psfrag{-1}[c][c][1][90]{\tiny -1}
  \psfrag{1}[c][c][1][90]{\tiny 1}
  \psfrag{2}[c][c]{\tiny 2}
  \psfrag{3}[c][c]{\tiny 3}
  \psfrag{0}[c][c][1]{\tiny 0}
  \psfrag{4}[c][c]{\tiny 4}
  \psfrag{10}[c][c]{\tiny 10}
  \psfrag{20}[c][c]{\tiny 20}
  \psfrag{30}[c][c]{\tiny 30}
  \psfrag{40}[c][c]{\tiny 40}
  \psfrag{50}[c][c]{\tiny 50}
 \psfrag{60}[c][c]{\tiny 60}
 \psfrag{80}[c][c]{\tiny 80}
  \psfrag{100}[c][c]{\tiny 100}
  \subfigure[][\label{fig:8a}]{\includegraphics[width=3.5cm,height=4cm]{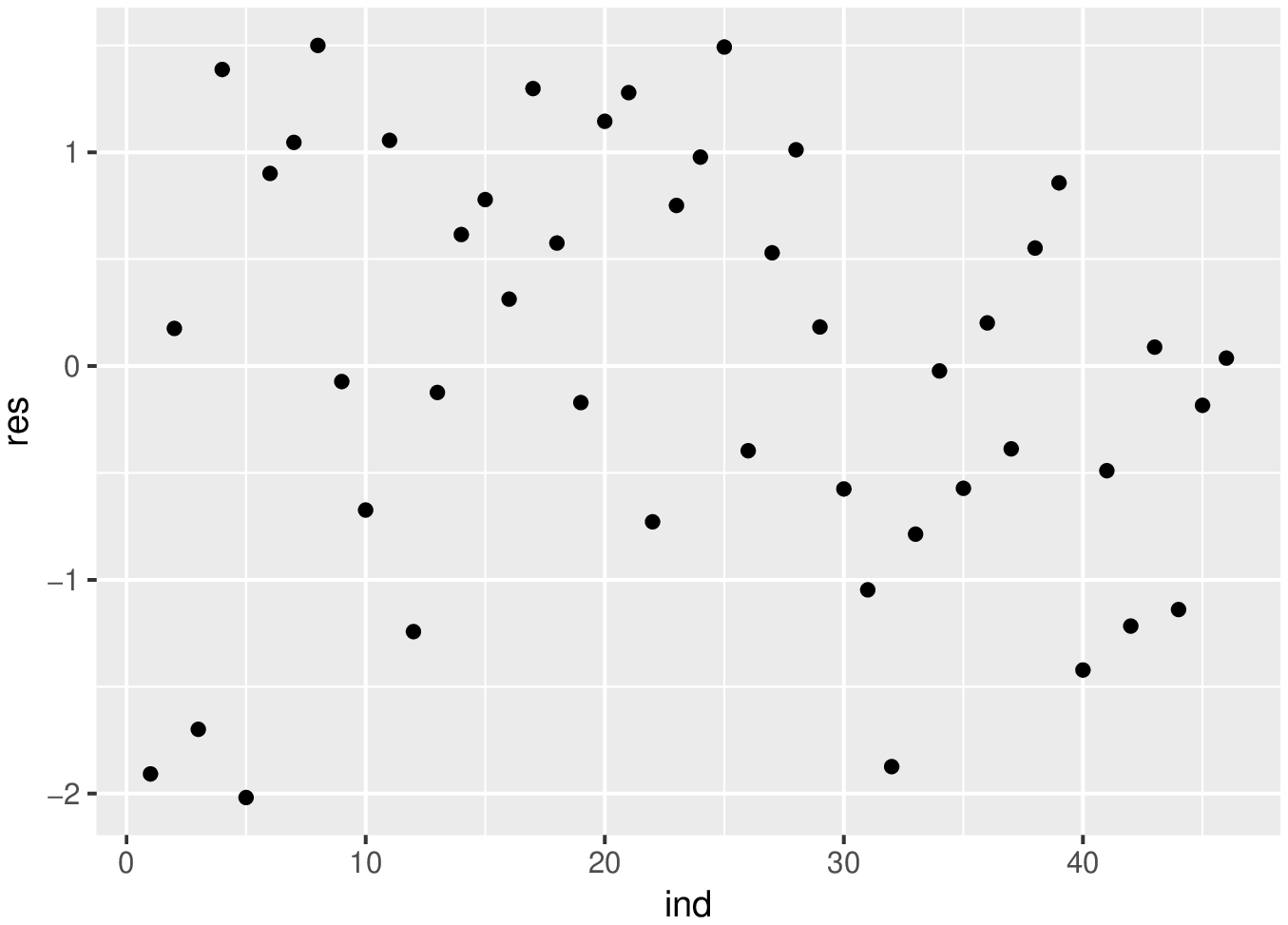}}\qquad
  \subfigure[][\label{fig:8b}]{\includegraphics[width=3.5cm,height=4cm]{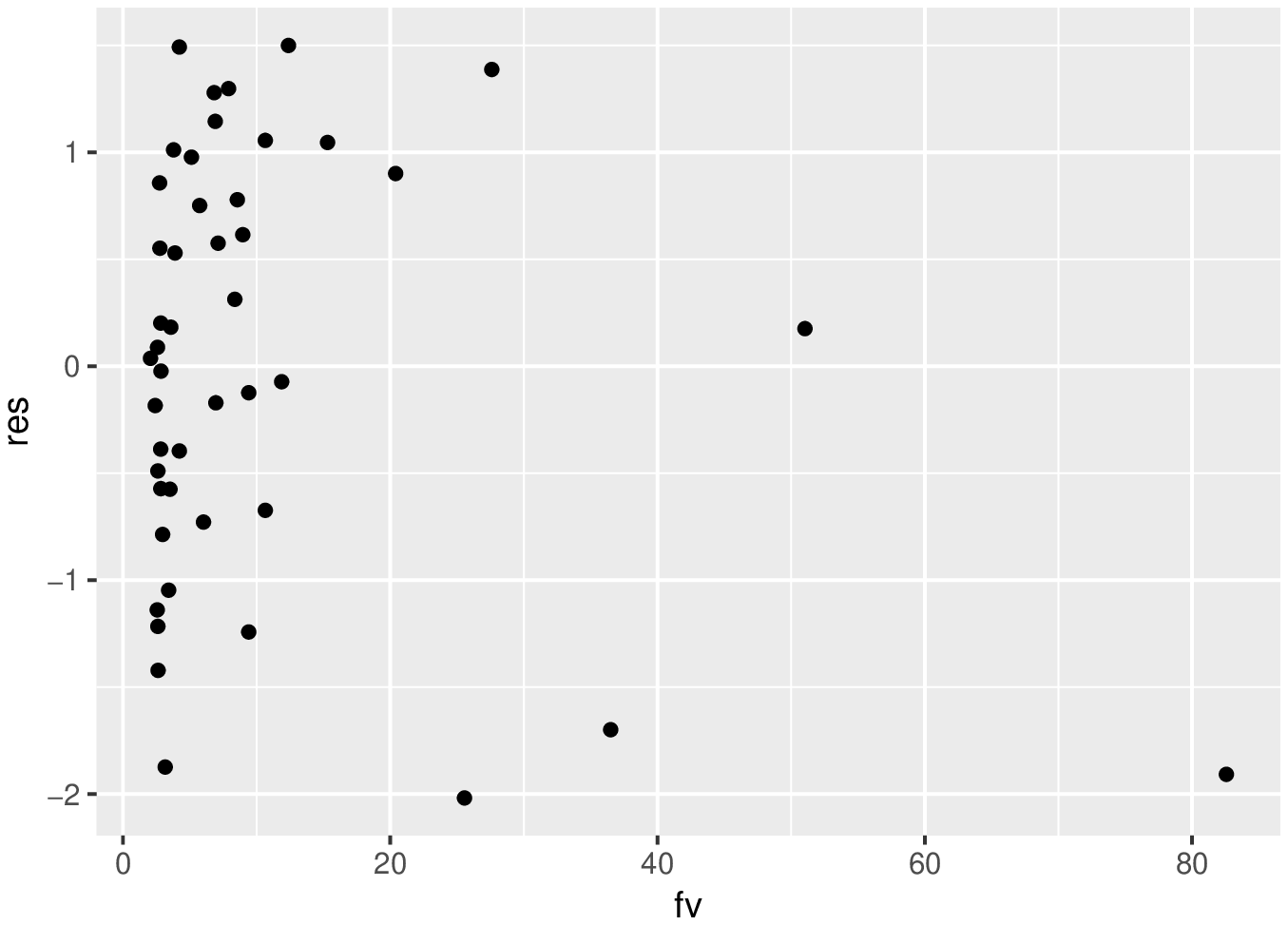}}
  \caption{Residuals plots.}
  \label{fig:8}
  \end{center}
  \end{figure}

\begin{figure}[ht]
  \begin{center}
  \psfrag{Cibeta}[c][c][0.8]{\tiny $\mathrm{C}_i$}
  \psfrag{Cialpha}[c][c][0.8]{\tiny $\mathrm{C}_i$}
  \psfrag{Citheta}[c][c][0.8]{\tiny $\mathrm{C}_i$}
  \psfrag{I}[c][c][0.8]{\tiny index}
  \psfrag{1.00}[c][c][0.8][90]{\tiny 1.00}
  \psfrag{0.75}[c][c][0.8][90]{\tiny 0.75}
  \psfrag{0.50}[c][c][0.8][90]{\tiny 0.50}
  \psfrag{0.25}[c][c][0.8][90]{\tiny 0.25}
  \psfrag{0.00}[c][c][0.8][90]{\tiny 0.00}
  \psfrag{0}[c][c]{\tiny 0}
  \psfrag{6}[r][l]{\tiny 6}
  \psfrag{1}[c][r]{\tiny 1}
  \psfrag{40}[c][c][0.8]{\tiny 40}
  \psfrag{10}[c][c][0.8]{\tiny 10}
  \psfrag{20}[c][c][0.8]{\tiny 20}
  \psfrag{30}[c][c][0.8]{\tiny 30}
  \subfigure[\label{fig:cia1}][$\bm \beta$]{\includegraphics[width=3.5cm,height=4cm]{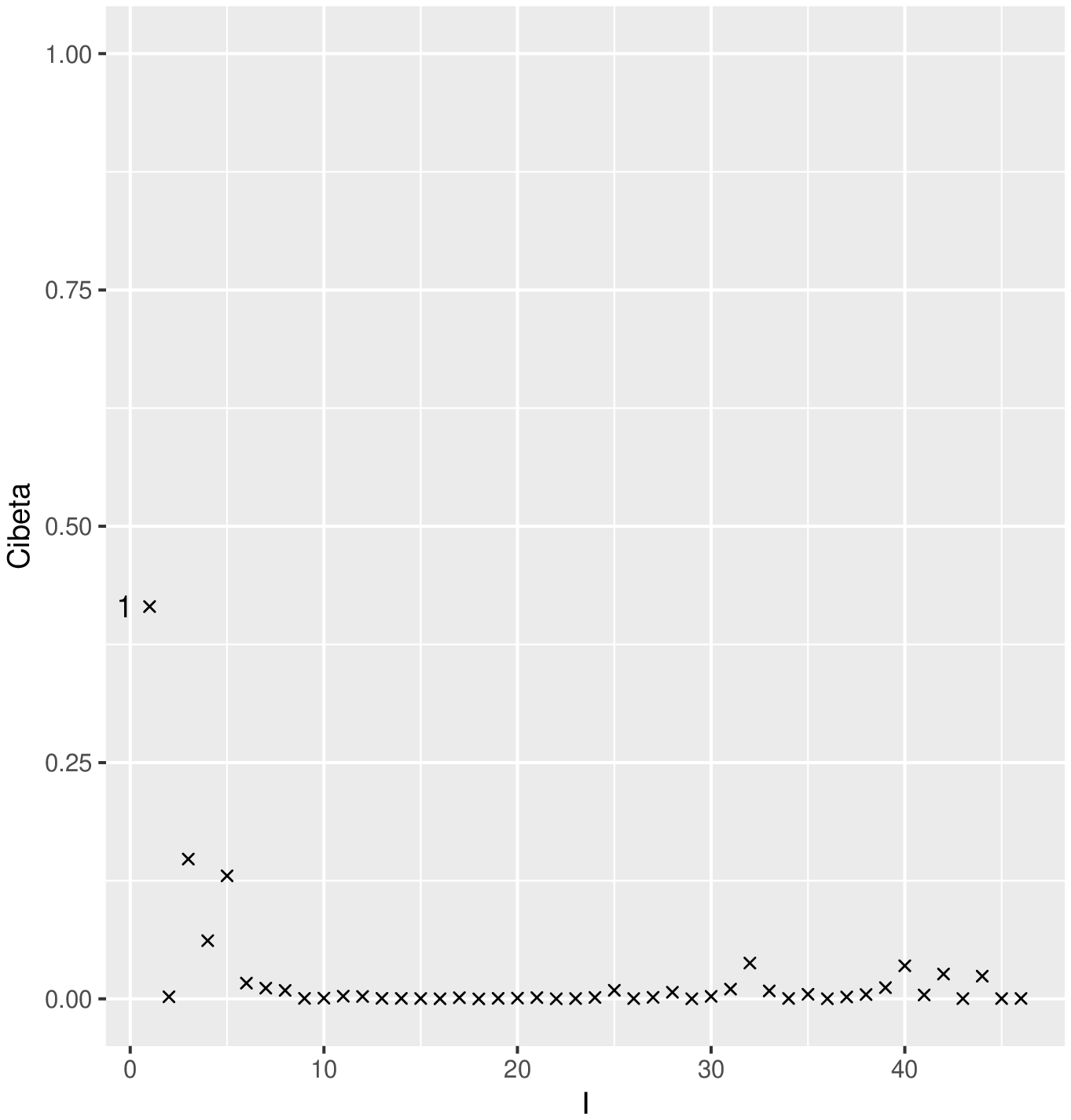}}\qquad
  \subfigure[\label{fig:cib1}][$\bm \alpha$]{\includegraphics[width=3.5cm,height=4cm]{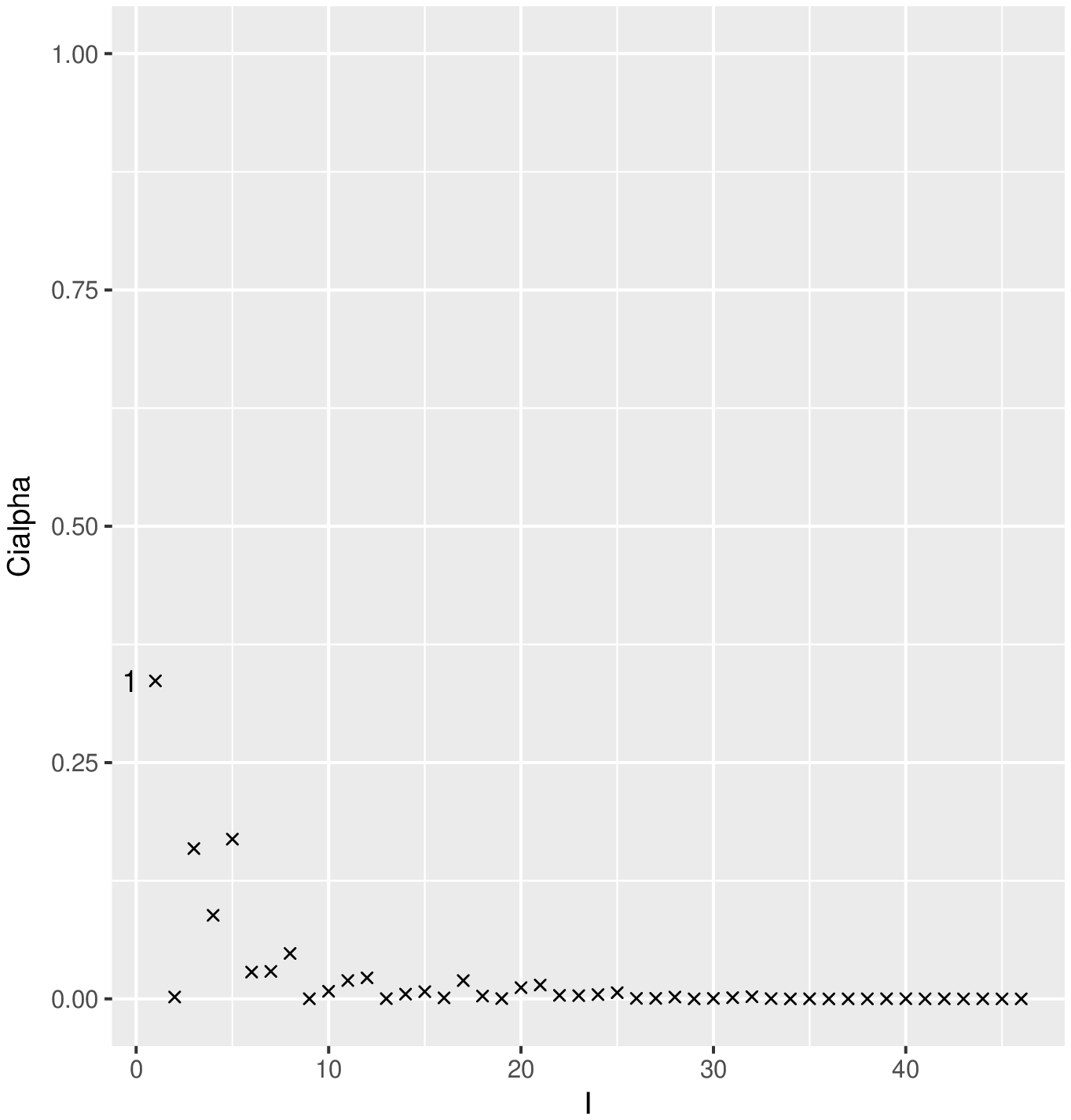}}\qquad
  \subfigure[\label{fig:cid1}][$\bm \theta$]{\includegraphics[width=3.5cm,height=4cm]{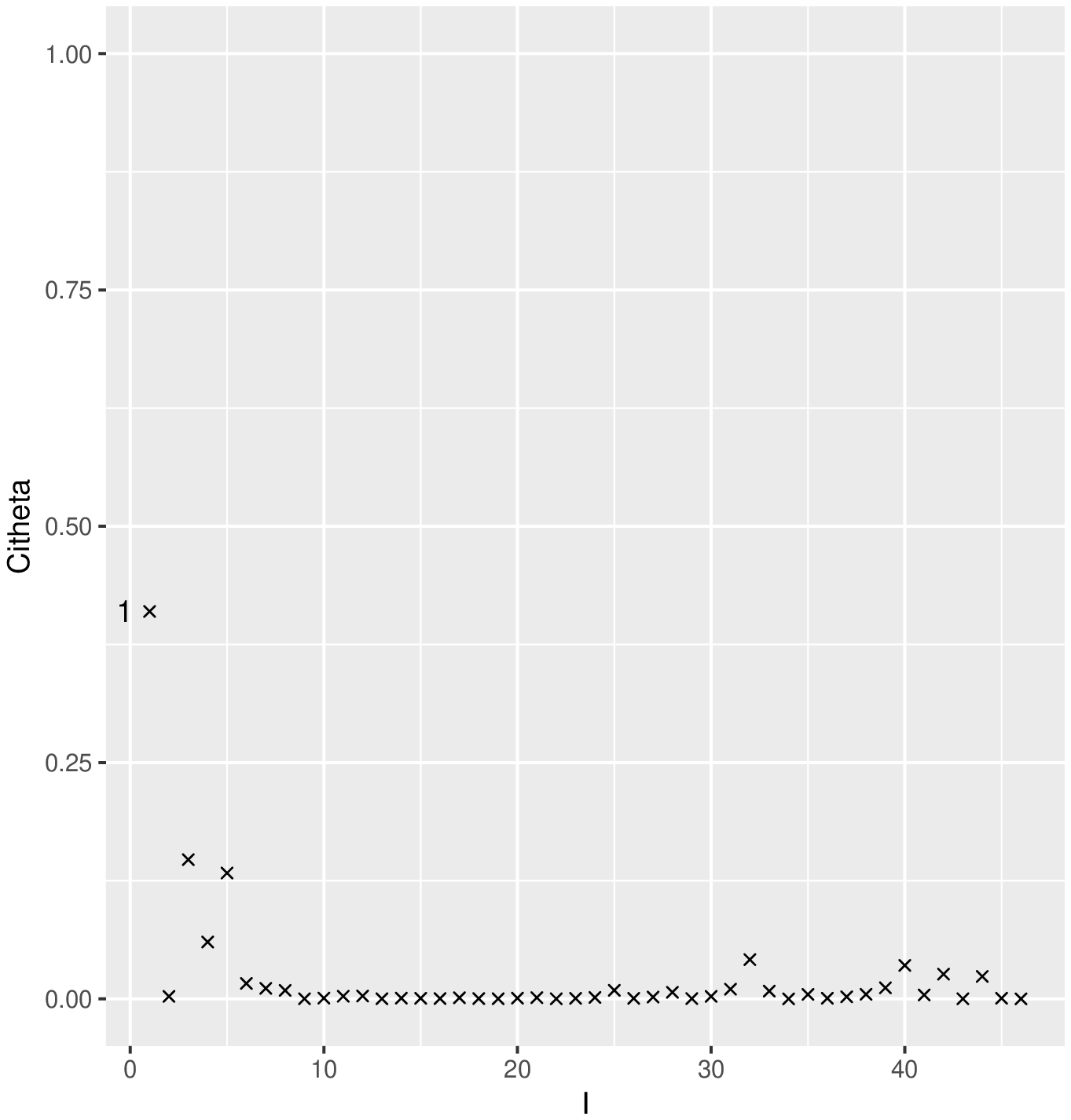}}
  \caption{$\mathrm{C}_i$'s plots.}
  \label{fig:ci1}
  \end{center}
  \end{figure}

\begin{figure}[ht]
  \begin{center}
  \psfrag{dmbeta}[c][c][0.8]{\tiny$|\mathbf{d}_\text{max}|$}
  \psfrag{dmalpha}[c][c][0.8]{\tiny $|\mathbf{d}_\text{max}|$}
  \psfrag{dmgamma}[c][c][0.8]{\tiny $|\mathbf{d}_\text{max}|$}
 \psfrag{dmtheta}[c][c][0.8]{\tiny $|\mathbf{d}_\text{max}|$}
  \psfrag{I}[c][c][0.8]{\tiny index}
  \psfrag{1.00}[c][c][0.8][90]{\tiny 1.00}
  \psfrag{0.75}[c][c][0.8][90]{\tiny 0.75}
  \psfrag{0.50}[c][c][0.8][90]{\tiny 0.50}
  \psfrag{0.25}[c][c][0.8][90]{\tiny 0.25}
  \psfrag{0.00}[c][c][0.8][90]{\tiny 0.00}
  \psfrag{0}[c][c]{\tiny 0}
  \psfrag{6}[r][l]{\tiny 6}
  \psfrag{7}[c][r]{\tiny 7}
  \psfrag{1}[c][c]{\tiny 1}
  \psfrag{40}[c][c][0.8]{\tiny 40}
  \psfrag{10}[c][c][0.8]{\tiny 10}
  \psfrag{20}[c][c][0.8]{\tiny 20}
  \psfrag{30}[c][c][0.8]{\tiny 30}
  \subfigure[\label{fig:lma1}][$\bm \beta$]{\includegraphics[width=3.5cm,height=4cm]{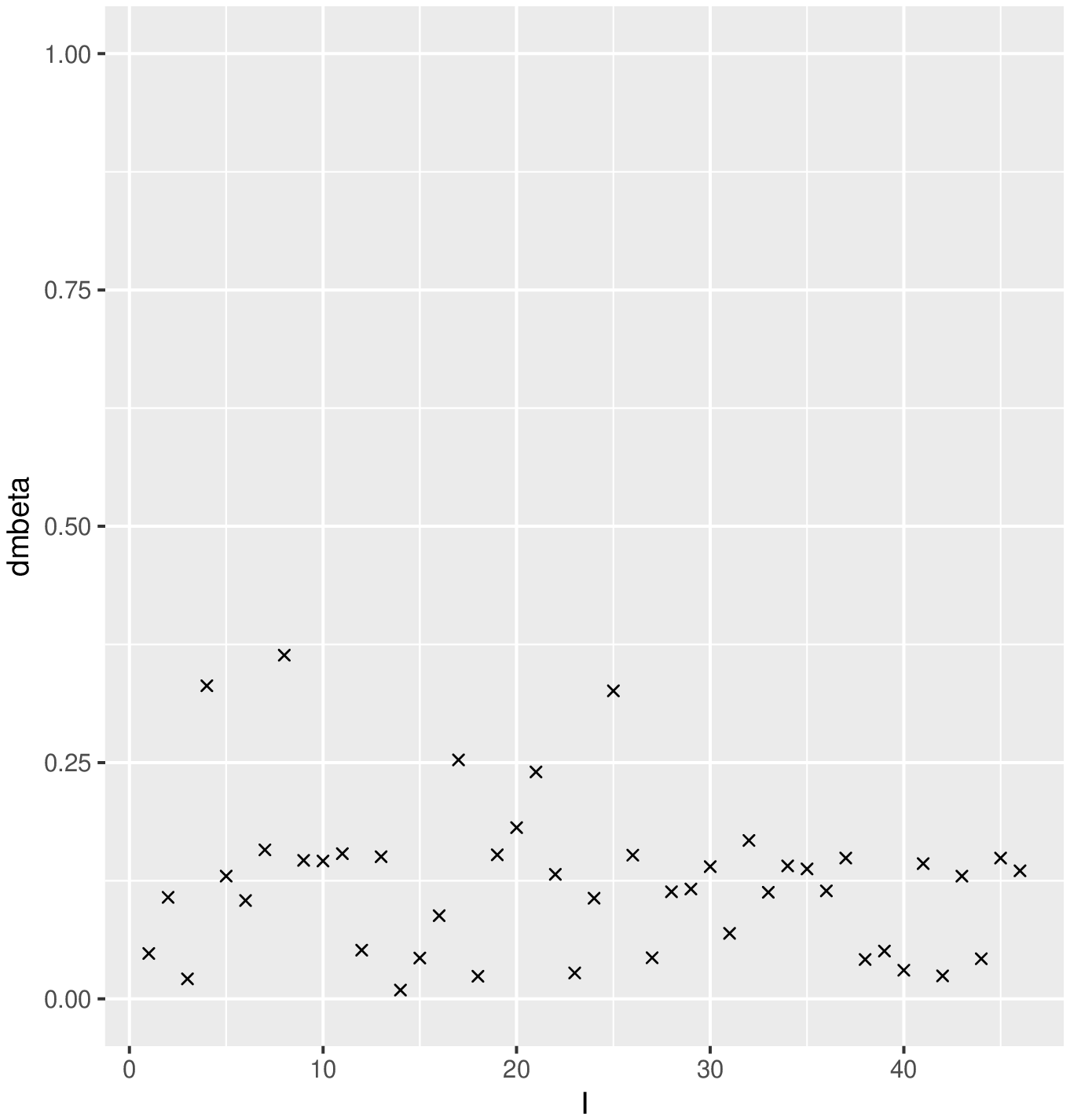}}\qquad
  \subfigure[\label{fig:lmb1}][$\bm \alpha$]{\includegraphics[width=3.5cm,height=4cm]{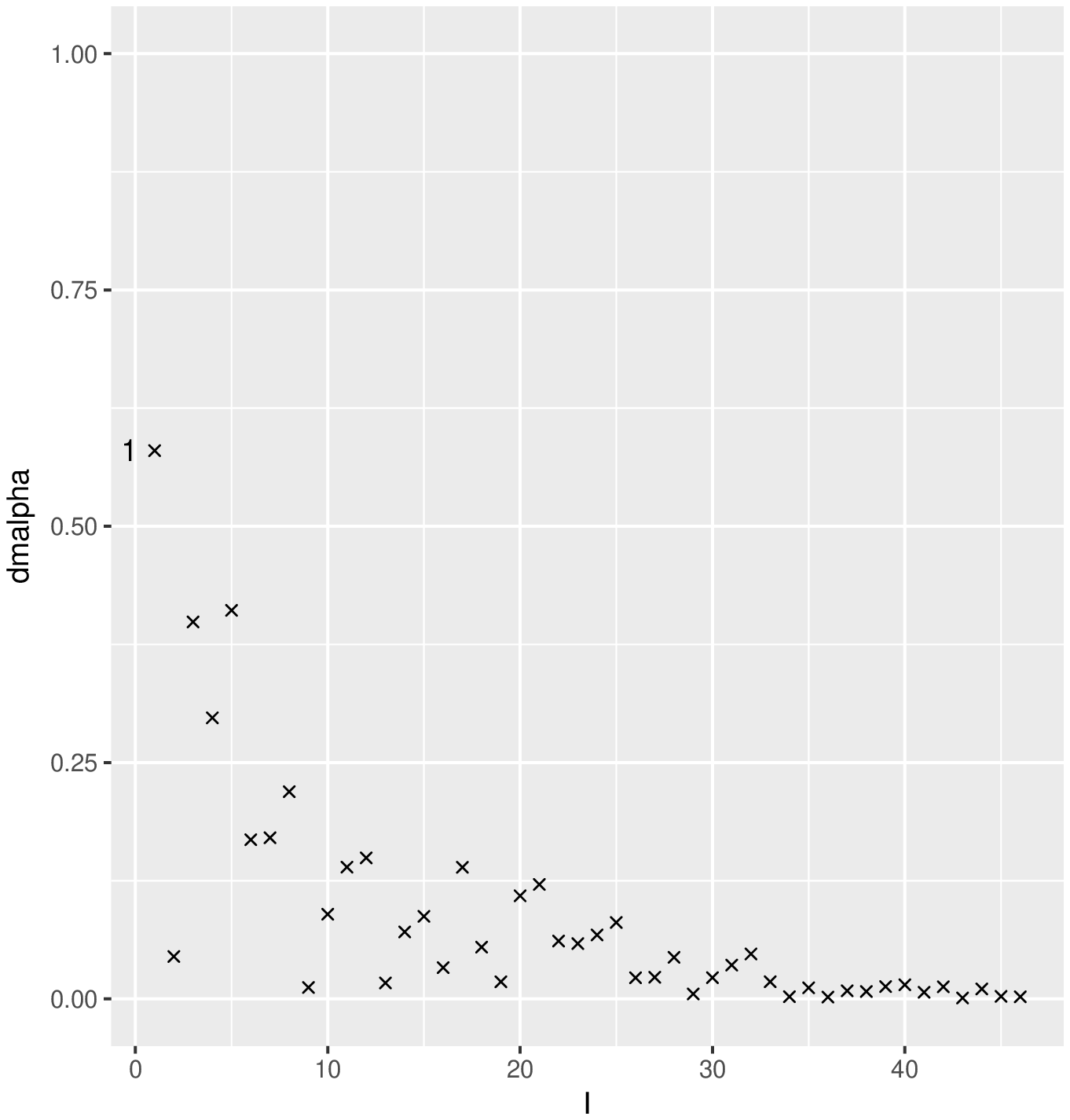}}\qquad
  \subfigure[\label{fig:lmd1}][$\bm \theta$]{\includegraphics[width=3.5cm,height=4cm]{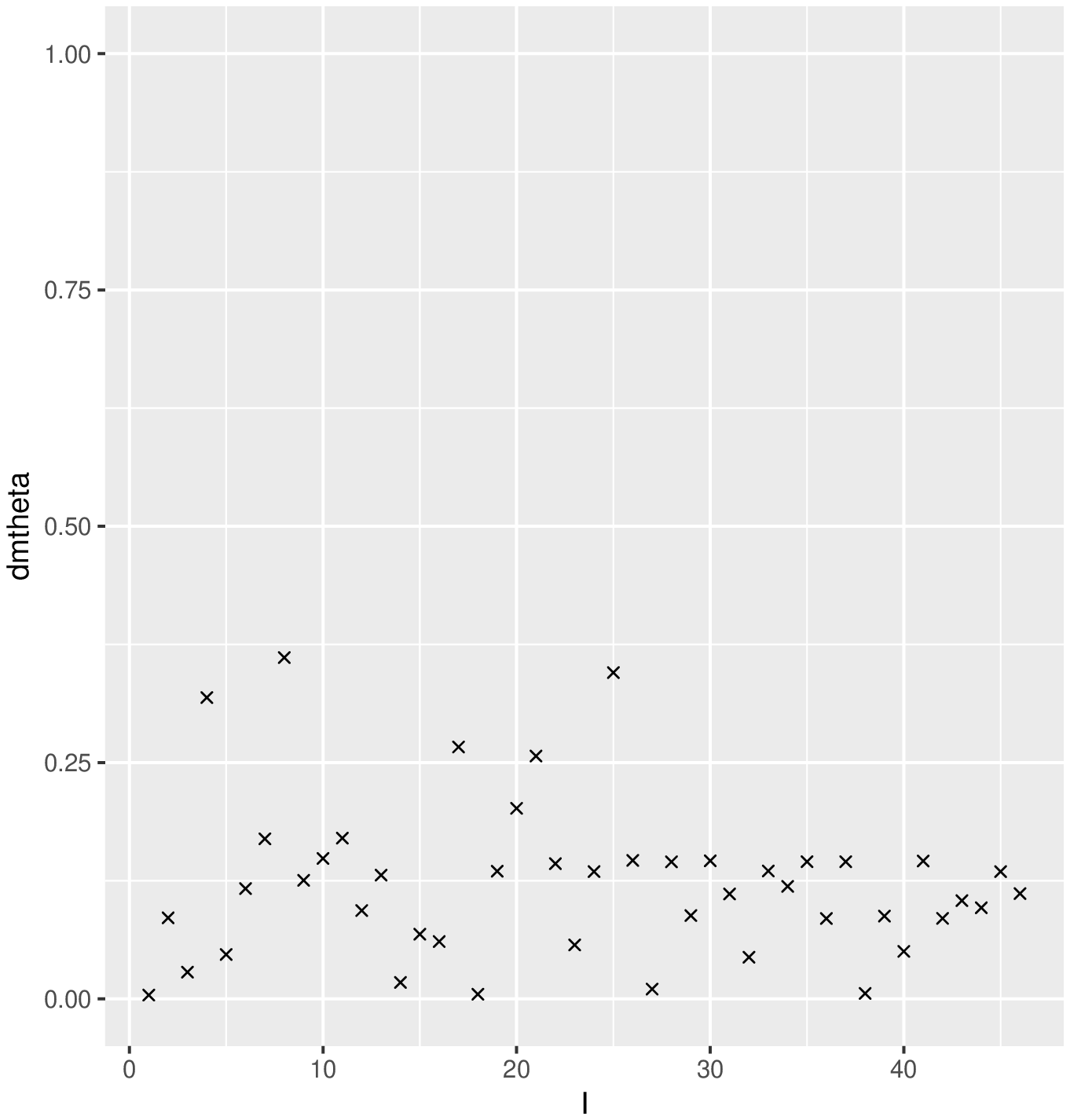}}
  \caption{Index plots of $|\mathbf{d}_\text{max}|$.}
  \label{fig:lm1}
  \end{center}
  \end{figure}

In this application, the data set considered is the biaxial fatigue data analyzed by \cite{rn:91}, \cite{lc:09}, \cite{Lemonte:2011aa}, \cite{vrc:12} and \cite{Lemonte:2016aa}. This data set has 46 observations and consists of lifes of a metal piece subjected to cyclic stretching and compressing, where the response variable $N$ denotes the number of cycles to failure of the metal specimen and the explanatory variable $W$ is the work per cycle ($mj/m^3$).  Here, we use the response variable on a reduced scale to facilitate the estimation procedure, i.e., $Y \times 100 \equiv N$. Therefore, we fit a BS nonlinear regression model 
\[
\mu_i = \beta_1 \textrm{e}^{\frac{\beta_2}{w_i}}, \quad i=1,\ldots,46,
\]
where $Y_i \sim \text{BS}(\mu_i,\sigma)$. The adjustment was performed using the function {\tt nlgamlss()} of the package {\tt gamlss.nl} whose objective  is to allow nonlinear fitting within a GAMLSS model. The estimates (standard errors) obtained were: $\hat \beta_1 = 1.276 (0.1719)$, $\hat \beta_2 = 47.954 (3.4402)$ and $\hat \sigma = 9.817 (2.049)$. Figure~\ref{fig:7} shows the QQ-plot with simulated envelope of quantile residuals and the scatter plot with the fitted model.  Figures~\ref{fig:7a} and \ref{fig:8} are used to verify the distribution assumption of the model and they do not show unusual features. Note that, in Figure~\ref{fig:7b}, the fitted line (proposed model) has a good agreement with the real data.
 Infuence diagnostics for the nonlinear BS regression model are presented in Figure \ref{fig:ci1}. Figures \ref{fig:ci1} and \ref{fig:lm1} displays index plots of $\mathrm{C}_i$ and $|\mathbf{d}_\text{max}|$, from where observation \#1 is detected as potentially influential. Changes are not detected when the observation is removed. 
We have that the mean of the number of cycles ($\times 100$) to failure of the metal specimen can be described by $\hat \mu(w_i) = 1.276\times \textrm{e}^{\frac{47.954}{w_i}}$.
\section{Concluding remarks} \label{sec5}
In this paper, we have introduced the ZABS regression model.  We have modeled the mean, the precision parameter and the probabilities of occurrences of zeros, through linear and/or nonlinear predictors, using appropriate link functions.  In addition, we have proposed randomized quantile residuals for our model. An iterative estimation procedure and its computational implementation have also been discussed. Moreover, we have considered a perturbation scheme which allows the identification of observations that exert unusual influence on the estimation process. In general, the results of the applications have shown the potentiality of proposed methodology. 
\section*{Acknowledgement}
We gratefully acknowledge grants from CAPES and CNPq (Brazil). We would like to thank Prof. Helton Saulo (UNB) and Prof. Jeremias Le\~ao (UFAM) for helpful comments on an earlier version of the manuscript.

\section*{Appendix}
\appendix
\section{Score vector, hessian and Fisher information matrix}\label{ap1}

First we will show how the elements of the score vector for this class of models were obtained. Let
\begin{eqnarray} \label{eq:sc}
\mathrm{U}_{\beta_j}=\sum_{i=1}^n d^{(i)}_\mu \, a_i \, \tilde{\mathrm{x}}_{ij}, \quad \mathrm{U}_{\alpha_R} =  \sum_{i=1}^n d^{(i)}_\sigma \, b_i \, \tilde{\mathrm{z}}_{iR}, \quad \text{and} \quad 
\mathrm{U}_{\gamma_J} =  \sum_{i=1}^n d^{(i)}_\nu \, c_i \, \tilde{\mathrm{w}}_{iJ}, 
\end{eqnarray}
where $\ell(\mu_i,\sigma_i, \nu_i)$ is defined in \eqref{ld} and for $j=1,\ldots,p$, $R=1,\ldots,q$ and $J=1,\ldots,r$ we have the following derived 
\begin{align} \label{eq:der1}
\begin{split}
d^{(i)}_\mu &=\underbrace{[1-\mathbb{I}(y_i=0)]}_{\kappa_i}\left\{
\underbrace{\frac{[\sigma_i+1]}{4\mu_i^2} -
\frac{\sigma_i^2}{4[\sigma_i+1]}\frac{1}{y_i} +
\frac{\sigma_i}{[\sigma_i+1]}\frac{1}{\left[y_i + \frac{\mu_i
\sigma_i}{\sigma_i+1}\right]}}_{y_i^{*}}-\underbrace{\frac{1}{2\mu_i}}_{\mu_i^{*}} \right\} = \kappa_i (y_i^{*} - \mu_i^{*}),  
\end{split}
\end{align}
\begin{align}\label{eq:der2}
\begin{split}
d^{(i)}_\sigma &=\underbrace{[1-\mathbb{I}(y_i=0)]}_{\kappa_i} \left\{ \underbrace{\frac{\mu_i}{[\sigma_i+1]^2}\frac{1}{\left[y_i + \frac{\mu_i
\sigma_i}{\{\sigma_i+1\}}\right]} -
\frac{1}{4\mu_i} y_i - \frac{\mu_i\sigma_i[\sigma_i+2]}{4[\sigma_i+1]^2}\frac{1}{y_i}}_{y_i^{\bullet}} + \underbrace{\frac{[\sigma_i+2]}{2[\sigma_i+1]}}_{-\sigma_i^{\bullet}}
\right\} = \kappa_i (y_i^{\bullet}-\sigma_i^{\bullet}),
\end{split}
\end{align}
and
\begin{align}\label{eq:der3}
\begin{split}
d^{(i)}_\nu &= \frac{1}{\nu_i}\,\mathbb{I}(y_i=0) - \frac{1}{1-\nu_i}\,[1-\mathbb{I}(y_i=0)] 
= \underbrace{\frac{1}{\nu_i[1-\nu_i]}\,\mathbb{I}(y_i=0)}_{y_i^{\circ}} - \underbrace{\frac{1}{[1-\nu_i]}}_{\nu_i^{\circ}} = (y_i^{\circ} - \nu_i^{\circ}),
\end{split}
\end{align}
where $a_i = \frac{1}{g_1'(\mu_i)} $, $b_i = \frac{1}{g_2'(\sigma_i)}$ and $c_i = \frac{1}{g_3'(\nu_i)}$. 
Therefore, using \eqref{eq:sc} and \eqref{eq:der1}-\eqref{eq:der3}, we can write the $(p+q+r)\times 1$ score vector $\mathbf{U}_{{\bm \theta}}$ in form 
$[\mathbf{U}_{{\bm \beta}}^\top, \mathbf{U}_{{\bm \alpha}}^\top,\mathbf{U}_{{\bm \gamma}}^\top]^\top$ with $\mathbf{U}_{{\bm \beta}} 
= \widetilde{\mathbf{X}}^\top\, \mathbf{A}_\kappa\,[\mathbf{y}^{*} - {\bm \mu}^{*}]$,
 $\mathbf{U}_{{\bm \alpha}} = \widetilde{\mathbf{Z}}^\top\, \mathbf{B}_\kappa\,[\mathbf{y}^{\bullet} - {\bm \sigma}^{\bullet}]$ and 
 $\mathbf{U}_{{\bm \gamma}}= \widetilde{\mathbf{W}}^\top\,\mathbf{C}\,[\mathbf{ y}^{\circ} - {\bm \nu}^{\circ} ]$, where
 $\mathbf{A}_\kappa = \kappa_i\, a_i\,\delta_{ij}^n$, $\mathbf{B}_\kappa = \kappa_i\, b_i\,\delta_{ij}^n$ and $\mathbf{C} = c_i\,\delta_{ij}^n$,
 with $\delta_{ij}^n$ being the Kronecker delta. Thus, $\mathbf{A}_\kappa$, $\mathbf{ B}_\kappa$ and $\mathbf{C}$
 are three $n \times n$ diagonal matrices. Moreover, we have to
  $\widetilde{\mathbf{X}} = \partial {\bm \eta}/\partial {\bm \beta}$, 
 $\widetilde{\mathbf{Z}} = \partial {\bm \tau}/\partial {\bm \alpha}$ and $\widetilde{\mathbf{W}} = \partial {\bm \xi}/\partial {\bm \gamma}$
 are derivative matrices that have ranks $p'$, $q$ and $r$, respectively, for all $\bm \beta$, $\bm \alpha$ and $\bm\gamma$. The hessian matrix is given by 
\begin{equation}
\arraycolsep=2.5pt\def\arraystretch{2.2}
\setlength{\dashlinegap}{2pt}
\mathbf{h}_{\bm \theta}= \left[
\begin{array}{ccc:c}
\widetilde{\mathbf{X}}^\top\, \mathbf{D}\, \widetilde{\mathbf{X}} && \widetilde{\mathbf{X}}^\top\, \mathbf{M}\, \widetilde{\mathbf{Z}} & \mathbf{0} \\ 
\widetilde{\mathbf{Z}}^\top\, \mathbf{M}\, \widetilde{\mathbf{X}}&& \widetilde{\mathbf{Z}}^\top\, \mathbf{E}\, \widetilde{\mathbf{Z}} & \mathbf{0}\\  \hdashline
 \mathbf{0} && \mathbf{0} & \widetilde{\mathbf{W}}^\top\,\mathbf{O}\, \widetilde{\mathbf{W}} 
\end{array}
\right].
\label{eqm15}
\end{equation}
The elements of the block matrix $\mathbf{h}_{\bm \theta}$ defined in \eqref{eqm15}, can be obtained by the partial derivatives. For $j, j'=1,\dots,p'$, $R, R'=1,\dots,q$ and $J, J'=1,\dots,r$,
we have
\begin{equation}
\begin{aligned}
\frac{\partial^2 \ell({\bm \theta}) }{\partial \beta_j \partial \beta_{j'}} &  = \sum_{i=1}^n \underbrace{\left[d_{\mu^2}^{(i)}\, a_i + d_{\mu}^{(i)} a'_i \right] a_i}_{d_i} \tilde{\mathrm{x}}_{ij} \tilde{\mathrm{x}}_{ik} , \quad
\frac{\partial^2 \ell({\bm \theta}) }{\partial \beta_j \partial \alpha_R}   = \sum_{i=1}^n \underbrace{d_{\mu \sigma}^{(i)}\, a_i\, b_i}_{m_i} \tilde{\mathrm{x}}_{ij} \tilde{\mathrm{z}}_{ik}, \quad
\frac{\partial^2 \ell({\bm \theta}) }{\partial \beta_j \partial \gamma_J} = 0, \\
\frac{\partial^2 \ell({\bm \theta}) }{\partial \alpha_R \partial \alpha_{R'}}  &= \sum_{i=1}^n \underbrace{\left[d_{\sigma^2}^{(i)}\, b_i + d_{\sigma}^{(i)} b'_i \right] b_i}_{e_i} \tilde{\mathrm{z}}_{ij}\tilde{\mathrm{z}}_{ik},   \quad
\frac{\partial^2 \ell({\bm \theta}) }{\partial \alpha_R \partial \gamma_J} = 0, \quad
\frac{\partial^2 \ell({\bm \theta}) }{\partial \gamma_J \partial \gamma_{J'}}   = \sum_{i=1}^n \underbrace{\left[d_{\nu^2}^{(i)}\, c_i + d_{\nu}^{(i)} c'_i \right] c_i}_{o_i} \tilde{\mathrm{w}}_{ij} \tilde{\mathrm{w}}_{ik}, 
\end{aligned}
\label{has}
\end{equation}
where
\begin{equation}
\begin{aligned}
d_{\mu^2}^{(i)} &=\kappa_i\,\left\{\frac{1}{2\mu_i^2} -\frac{\sigma_i^2}{[\sigma_i y_i + y_i + \sigma_i \mu_i]^2} - {\frac {y_i[\sigma_i+1]}{2{\mu_i}^{3}}} \right\}, \quad
d_{\mu \nu}^{(i)} = 0,  \\ 
d_{\mu\sigma}^{(i)} &=\kappa_i\,\left\{\frac{y_i}{[\sigma_i y_i + y_i + \sigma_i \mu_i]^2}  + \frac{y_i}{4\mu^2_i} -\frac{\sigma_i[\sigma_i+2]}{4[\sigma_i+1]^2y_i} \right\}, \quad
d_{\sigma \nu}^{(i)} = 0,\\
d_{\sigma^2}^{(i)} &=\kappa_i\,\left\{\frac{1}{2[\sigma_i+1]^2} - \frac{[y_i+\mu_i]^2}{[\sigma_i y_i + y_i +\sigma_i \mu_i]^2} -\frac{\mu_i}{2[\sigma_i+1]^3 y_i} \right\}, \quad
d_{\nu^2}^{(i)} = \frac{2\nu_i - 1}{\nu_i^2[1-\nu_i]^2}\,\mathbb{I}(y_i=0) - \frac{1}{[1-\nu_i]^2}.
\end{aligned}
\label{sder}
\end{equation}
Using the quantities presented in \eqref{has} and \eqref{sder}  we obtain the following matrices $\mathbf{h}_{\bm \beta} = \widetilde{\mathbf{X}}^\top\, \mathbf{D}\, \widetilde{\mathbf{X}} $,
$\mathbf{h}_{{\bm \beta}{\bm \alpha}} = \widetilde{\mathbf{X}}^\top\, \mathbf{M}\, \widetilde{\mathbf{Z}}$, $\mathbf{h}_{{\bm \alpha}}   
= \widetilde{\mathbf{Z}}^\top\, \mathbf{E}\, \widetilde{\mathbf{Z}}$ and $\mathbf{h}_{{\bm  \gamma}} = \widetilde{\mathbf{W}}^\top \mathbf{O}\, \widetilde{\mathbf{W}}$,
with $\mathbf{M} = m_i\, \delta_{ij}^n$, $\mathbf{D} = d_i\, \delta_{ij}^n$, $\mathbf{E} = e_i\, \delta_{ij}^n$ and $\mathbf{O} = o_i\, \delta_{ij}^n$. Thus, the corresponding Fisher information matrix is given by
\begin{equation}
\arraycolsep=2.5pt\def\arraystretch{2.2}
\setlength{\dashlinegap}{2pt}
\mathbf{i}_{\bm \theta}= \left[
\begin{array}{ccc:c}
\widetilde{\mathbf{X}}^\top\, \mathbf{V}\, \widetilde{\mathbf{X}} & &\widetilde{\mathbf{X}}^\top\, \mathbf{S}\, \widetilde{\mathbf{Z}} & \mathbf{0} \\ 
\widetilde{\mathbf{Z}}^\top\,  \mathbf{S}\, \widetilde{\mathbf{X}}&& \widetilde{\mathbf{Z}}^\top\,  \mathbf{U}\, \widetilde{\mathbf{Z}} & \mathbf{0}\\  \hdashline
 \mathbf{0} && \mathbf{0} & \widetilde{{\bm W}}^\top\, \mathbf{Q}\, \widetilde{\mathbf{W}} 
\end{array}
\right],
\label{inf}
\end{equation}
where $\mathbf{V} = \mathcal{E}_{d_i}\,\sigma_{ij}^n = \diag\{\mathcal{E}_{d_1},\ldots,\mathcal{E}_{d_n}\}$, 
$\mathbf{S} = \mathcal{E}_{m_i}\,\sigma_{ij}^n = \diag\{\mathcal{E}_{m_1},\ldots,\mathcal{E}_{m_n}\}$, 
$\mathbf{U} = \mathcal{E}_{e_i}\,\sigma_{ij}^n = \diag\{\mathcal{E}_{e_1},\ldots,\mathcal{E}_{e_n}\}$,
$\mathbf{Q} = \mathcal{E}_{o_i}\,\sigma_{ij}^n = \diag\{\mathcal{E}_{o_1},\ldots,\mathcal{E}_{o_n}\}$, with 
$\mathcal{E}_{d_i} = -\mathbb{E}[d_i] = [1-\nu_i]\left\{\displaystyle\frac{\sigma_i}{2\mu_i^2} + \displaystyle\frac{\sigma_i^2}{[\sigma_i+1]^2}\, \lambda_i \right\}\, a_i^2$,
$\mathcal{E}_{m_i} = -\mathbb{E}[m_i] = [1-\nu_i]\left\{\displaystyle\frac{1}{2\mu_i[\sigma_i+1]} + \displaystyle\frac{\sigma_i \mu_i}{[\sigma_i+1]^3} \,\lambda_i \right\}\, a_i\,b_i$,
$\mathcal{E}_{e_i} = -\mathbb{E}[e_i] = [1-\nu_i]\left\{\displaystyle\frac{[\sigma_i^2+3\sigma_i+1] }{2\sigma_i^2[\sigma_i+1]^2} +\displaystyle \frac{\mu_i^2}{[\sigma_i+1]^4}\,\lambda_i \right\}\, b_i^2$,
$\mathcal{E}_{o_i} = -\mathbb{E}[o_i] = \displaystyle\frac{1}{\nu_i[1-\nu_i]} $ and $\lambda_i = \displaystyle \int_{0}^{\infty}  \frac{\sqrt{\sigma_i+1} \exp\{\sigma_i/2\}}{4\sqrt{\pi
 \mu_i}y_i^{3/2}} \left[y_i+ \frac{\sigma_i\mu_i}{\sigma_i+1}\right]^{-1}
 \exp\left(-\frac{\sigma_i}{4}\left[ {\frac{[\sigma_i+1]y_i}{\sigma_i\mu_i}}+{\frac
 {\sigma_i\mu_i}{[\sigma_i+1] y_i}} \right]\right)\text{d}y_i.$
 The integral $\lambda_i$ can be calculated numerically using the {\tt integrate} function
 of the {\tt R} software. The inverse Fisher information matrix, defined in \eqref{inf}, is given by
 \begin{equation}
\def\arraystretch{2.2}
\mathbf{i}_{\bm \theta}^{-1}= \left[
\begin{array}{cc:c}
[\widetilde{\mathbf{X}}^\top\, \mathbf{W}_1 \widetilde{\mathbf{X}}]^{-1} & -[\widetilde{\mathbf{X}}^\top\, \mathbf{W}_1 \widetilde{\mathbf{X}}]^{-1} \widetilde{\mathbf{X}}^\top\, \mathbf{S} \widetilde{\mathbf{Z}} [\widetilde{\mathbf{Z}}^\top\,  \mathbf{U} \widetilde{\mathbf{Z}}]^{-1}  & \mathbf{0} \\ 
-[[\widetilde{\mathbf{X}}^\top\, \mathbf{W}_1 \widetilde{\mathbf{X}}]^{-1} \widetilde{\mathbf{X}}^\top\, \mathbf{S} \widetilde{\mathbf{Z}} [\widetilde{\mathbf{Z}}^\top\,  \mathbf{U} \widetilde{\mathbf{Z}}]^{-1}]^{\top}& [\widetilde{\mathbf{Z}}^\top\,  \mathbf{W}_2 \widetilde{\mathbf{Z}}]^{-1} & \mathbf{0}\\  \hdashline
 \mathbf{0} & \mathbf{0} & [\widetilde{\mathbf{W}}^\top \mathbf{Q} \widetilde{\mathbf{W}}]^{-1} 
\end{array}
\right],
\label{invF}
\end{equation}
 where $\mathbf{W}_1 = \mathbf{V} - \mathbf{S}\widetilde{\mathbf{Z}} [\widetilde{\mathbf{Z}}^\top\,  \mathbf{U} \widetilde{\mathbf{Z}} ]^{-1}\widetilde{\mathbf{Z}}^\top\, \mathbf{S}$ and 
 $\mathbf{W}_2 = \mathbf{U} - \mathbf{S}\widetilde{\mathbf{X}} [\widetilde{\mathbf{X}}^\top\,\mathbf{V}\, \widetilde{\mathbf{X}} ]^{-1}\widetilde{\mathbf{X}}^\top\, \mathbf{S}$.
Using the augmented matrices, $\breve{\mathbf{W}}$, $\breve{\mathbf{X}}$ and $\breve{\mathbf{D}}$,
 \begin{equation*}
 \small
 \arraycolsep=1.2pt\def\arraystretch{1.7}
\breve{\mathbf{W}}_{3n,3n}= \left[
\begin{array}{cc:c}
 \mathbf{V}  &  \mathbf{S} & \mathbf{0} \\ 
 \mathbf{S} &   \mathbf{U}  & \mathbf{0}\\  \hdashline
 \mathbf{0} & \mathbf{0} &  \mathbf{Q} 
\end{array}
\right], 
\breve{\mathbf{X}}_{3n,p'+q+r}= \left[
\begin{array}{cc:c}
 \widetilde{\mathbf{X}}  &  \mathbf{0} & \mathbf{0} \\ 
 \mathbf{0} &   \widetilde{\mathbf{Z}}  & \mathbf{0}\\  \hdashline
 \mathbf{0} & \mathbf{0} &  \widetilde{\mathbf{W}}
\end{array}
\right], \,\text{and}\, \breve{\mathbf{D}}_{3n,3n}=\left[\begin{array}{cc:c}
  \mathbf{A}^{(m)}& \mathbf{0} & \mathbf{0}  \\
\mathbf{0} & \mathbf{B}^{(m)} &  \mathbf{0} \\ \hdashline
 \mathbf{0} & \mathbf{0}& \mathbf{C}^{(m)} \\
\end{array} \right]
\end{equation*}
the Fisher information matrix can be rewritten as $\mathbf{i}_{\bm \theta} = \breve{\mathbf{X}}^\top\, \breve{\mathbf{W}}\,\breve{\mathbf{X}}$
and its inverse as $\mathbf{i}_{\bm \theta}^{-1} =[\breve{\mathbf{X}}^\top\, \breve{\mathbf{W}}\,\breve{\mathbf{X}}]^{-1}$. By using the Fisher scoring iterative
procedure, the corresponding estimation algorithm for ${\bm \theta} = [{\bm \beta}^\top, {\bm \alpha}^\top, {\bm \gamma}^\top ]^\top$ is given by
$
{\bm \theta}^{(m+1)} = {\bm \theta}^{(m)} + \mathbf{i}_{\bm \theta}^{-1} \mathbf{U}_{\bm \theta}^{(m)} 
 = [\breve{\mathbf{X}}^\top \breve{\mathbf{W}}^{(m)} \breve{\mathbf{X}}]^{-1} \breve{\mathbf{X}}^\top \breve{\mathbf{W}}^{(m)} \tilde{\mathbf{z}}^{(m)}
$
where $\tilde{\mathbf{z}}^{(m)} = \breve{\mathbf{X}} {\bm \theta}^{(m)} + \{\breve{\mathbf{W}}^{(m)}\}^{-1}\breve{\mathbf{D}}^{(m)} \breve{\mathbf{z}}^{(m)} $
and $\breve{\mathbf{z}}^{(m)} = [[\mathbf{y}^{*} - {\bm \mu}^{*}]^{(m)\top}, [\mathbf{y}^{\bullet} - {\bm \delta}^{\bullet}]^{(m)\top}, [\mathbf{y}^{\circ} - {\bm \nu}^{\circ} ]^{(m)\top}  ]^\top$.

%

\end{document}